\documentclass{article}
\usepackage{techart}
\pdfoutput=1

\renewcommand\vec[1]{\boldsymbol{#1}}
\renewcommand\mat[1]{\boldsymbol{#1}}

\newcommand\x{x}
\newcommand\y{y}
\newcommand\z{z}

\newcommand\Y{Y}

\newcommand\vx{\vec{\x}}
\newcommand\vy{\vec{\y}}
\newcommand\vY{\vec{\Y}}
\newcommand\vz{\vec{\z}}

\newcommand\mX{\mat{X}}

\newcommand\vfx{\tilde{\vx}}
\newcommand\vfy{\tilde{\vy}}
\newcommand\fY{\widetilde{\vY}}

\newcommand\fM{\mathcal{M}}

\newcommand\mR{\mat{R}}
\newcommand\mQ{\mat{Q}}
\newcommand\mH{\mat{H}}
\newcommand\mW{\mat{W}}

\newcommand\vbeta{\vec{\beta}}
\newcommand\vtheta{\vec{\theta}}
\newcommand\vvartheta{\vec{\vartheta}}
\newcommand\vTheta{\vec{\Theta}}

\newcommand\vmu{\vec{\mu}}
\newcommand\mSigma{\mat{\Sigma}}

\newcommand\verr{\vec{\epsilon}}
\newcommand\verra{\verr\ped{a}}
\newcommand\verrb{\verr\ped{b}}

\newcommand\vza{\vz\ped{a}}
\newcommand\vzb{\vz\ped{b}}
\newcommand\vvarthetaa{\vvartheta\ped{a}}
\newcommand\vvarthetab{\vvartheta\ped{b}}

\newcommand\gauss{N}

\begin{document}
\title{Modeling Gaussian Random Fields by Anchored Inversion\\
and Monte Carlo Sampling}
\author{Zepu Zhang}
\date{July 14, 2009}
\maketitle

\barefoot{%
Copyright \copyright\@ 2009 Zepu Zhang zepu.zhang@gmail.com. All rights reserved.\\
Title and abstract revised on July 24, 2009.\\[1mm]
Parts of this article are written in a way that is not easy to
understand. However, this article is a milestone in the development of
the ``anchored inversion'' methodology.
It is publicized for historical reasons.\\[1mm]
See the commentary arXiv:1104.0475 [stat.AP] at
\url{http://arxiv.org/abs/1104.0475}.\\
Additional information on this work can be found at
\url{http://www.AnchoredInversion.info/}}
\vspace*{-1cm}


\begin{abstract}
It is common and convenient to treat distributed physical parameters
as Gaussian random fields and model them in an ``inverse procedure''
using measurements of various properties of the fields.
This article presents a general method for this problem based on
a flexible parameterization device called ``anchors'',
which captures local or global features of the fields.
A classification of all relevant data into two categories
closely cooperates with the anchor concept to enable systematic use of
datasets of different sources and disciplinary natures.
In particular, nonlinearity in the ``forward models'' is handled automatically.
Treatment of measurement and model errors is systematic and integral in
the method; however the method is also suitable in the usual setting
where one does not have reliable information about these errors.
Compared to a state-space approach,
the anchor parameterization renders the task in a
parameter space of radically reduced dimension;
consequently, easier and more rigorous
statistical inference, interpretation, and sampling are possible.
A procedure for deriving the posterior distribution of model
parameters is presented.
Based on Monte Carlo sampling and normal mixture approximation to
high-dimensional densities,
the procedure has generality and efficiency features
that provide a basis for
practical implementations of this computationally demanding inverse procedure.
We emphasize distinguishing features of the method compared to
state-space approaches and optimization-based ideas.
Connections with existing methods in stochastic hydrogeology are discussed.
The work is illustrated by a one-dimensional synthetic problem.


\end{abstract}

\section{Introduction}
\label{sec:intro}

The term ``inverse problems'' is extremely broad,
and the literature is vast.
Considered in this paper is one type of inverse problems,
which concern inferring a spatially distributed, highly heterogeneous
attribute from a finite number of relevant data.
``Spatially distributed'', or ``point referenced'',
means that in principle the attribute is a function of the spatial
location (geometric point).
Relevant data include measurements of the attribute in question
or covariates,
and measurements of responses of the attribute field to certain forcings.
Such problems arise in many situations, as shown in the following
examples.

\emph{Example 1}.
\citet{Yeh:1986:RPI} reviews techniques for the groundwater
inverse problem.
Groundwater flow and transport are controlled by
hydraulic conductivity or transmissivity of the medium.
These spatially distributed attributes can be measured,
with high uncertainty, at a small number of locations
but can not be resolved throughout the spatial domain of interest,
However,
one can make observations of natural or experimental flow and
transport processes to get head (water level), flux (water flow),
and tracer concentrations at selected locations and times.
The flow and transport processes are governed by partial differential
equations.
One's task is to infer the conductivity or transmissivity field
given the above observations.

\emph{Example 2}.
\citet{Bube:1983:ODI} study a 1-D problem for exploration seismology.
In this setting,
an impulsive or vibrating load applied at the ground surface launches
elastic waves into the earth's interior.
Part of the wave energy is reflected and reaches back the ground
surface, where it is monitored at many times.
Mechanic attributes of the elastic medium that control wave propagation
include density and Lam\'e constants;
``effects'' of the wave propagation include pressure and particle
velocity, which are measured at the ground surface.
The wave propagation is described by a hyperbolic equation system.
The task is to recover the mechanic attributes, or
transformations thereof, using the ground surface measurements.

\emph{Example 3}.
\citet{Newsam:1988:IPA} investigate the problem of deducing the spatial
distribution of sources (including sinks) of CO$_2$ on the surface of the globe,
given observations of surface concentrations of this non-reactive gas.
The source distribution is linked to concentration by a transport model,
governed by a diffusion equation.
Although the transport model describes CO$_2$ concentration in a spherical shell
12\unit{km} thick around the earth,
all available concentration measurements are at the earth's surface, with time series
records at some 20 locations around the globe.

In these examples,
the attribute of interest
(hydraulic properties of flow media, mechanic properties of elastic
media, and source/sink of gas)
is connected with the measurements
(head and flux, tracer concentration; pressure, particle velocity;
and CO$_2$ concentration) by a ``forward process''
(subsurface flow and transport, wave propagation, and atmospheric transport),
which is usually embodied
in a numerical model due to the spatial heterogeneity of the input
attribute field.
The attribute of interest is a function (or functional) of the spatial location,
hence is of infinite dimension, therefore the inverse problem is
ill-posed, or under-determined.
For most practical purposes, we assume the field is discretized into a
finite vector corresponding to a numerical grid.
As computational power advances,
people try to use finer and finer numerical grids in order to increase
the reliability of the forward model, as an approximation to the actual
process;
as a result, the size of the model grid still far exceeds the number of
measurements, hence the problem of inferring the attribute field
(vector) remains seriously ill-posed.

Because of the ill-posedness,
infinitely many configurations of the attribute field,
when fed into the forward model,
may produce equally good match to the observations,
yet some of these configurations may well contain features that are
physically unrealistic.
Perhaps the most commonly used solution to this problem is
``regularization''
\citep{Tikhonov:1977:SIP, Tenorio:2001:SRI},
which imposes subjective requirements on the structure (such as
smoothness) of the attribute field.
Usually, a ``model performance'' term,
which indicates how closely a particular solution of the attribute field
reproduces the measurements,
is optimized in search for attribute fields that achieve
a required level of performance, under the constraint of a
regularization term.
This approach is exemplified for groundwater inverse problems
by \citet{Doherty:2003:GWM, Gomez-Hernanez:1997:SST1,
Kowalsky:2004:EFP}.

In this class of inverse problems,
regularization is essentially always used,
explicitly or not,
in the form of certain assumptions on the spatial structure,
because brute force search for a ``good''
attribute field is both hopeless and meaningless.

Another response to the ill-posedness of the problem is that it has
become more or less a consensus to take a nondeterministic perspective,
treating the attribute field as a stochastic process
\citep{Kitanidis:1986:PUE, Rubin:2003:ASH}.
From this angle, a Bayesian interpretation is convenient and Monte Carlo
sampling methods are indispensable \citep{Sambridge:2002:MCM,
Robert:2005:MCS}.

Besides ill-posedness,
another main difficulty in this problem is nonlinearity in the forward
model with respect to the attribute in question.
The case with a linear forward model is largely solved by the Kalman
filter \citep[see an introduction by][]{Welch:1995:IKF}.
However, the vast majority of the forward models are nonlinear.
The Kalman filter has been extended to cope with nonlinearity by
``linearization'' \citep{Welch:1995:IKF}.
This strategy in hydrogeology is represented by the ``quasi-linear''
method of \citet{Kitanidis:1995:QLG}.
A direct attack at nonlinearity is the ensemble Kalman filter
\citep{Evensen:2003:EKF} and variants.

Most of the works mentioned above take a state-space approach,
which treats the attribute vector on the numerical model grid as the
model parameter vector, and outputs values of this vector (\ie
realizations of the attribute field) as the product of the inversion.
It is generally desired to provide an ensemble of realizations,
the spread of which present a measure of the uncertainty in the
solution. Often, the algorithm for generating each ensemble member is
identical, hence the computational cost is proportional to the size of the
ensemble.

This paper proposes a new statistical approach to this inverse problem.
The idea treats the unknown attribute field as a Gaussian process and
uses a basic property of the Gaussian process to achieve
parsimonious yet flexible parameterization of the field.
In no attempt to make a comprehensive assessment at this point,
we mention several distinguishing features of the proposed method,
which shall be called ``anchored inversion'' in this paper.

The method is not a state-space approach.
It describes the field with a relatively small number of parameters, say
several tens; this parameterization is very flexible,
and in principle is separate from the resolution of the numerical model grid.
This dramatic reduction in the parameter dimensionality
has advantages in parameter inference, sampling,
as well as statistical interpretation of the distributions of the model
parameters and field realizations.
The ill-posedness issue of this relatively short parameter
vector is much milder or even eliminated.

Result of anchored inversion is some representation of the
posterior distribution of the model parameters.
Generation of field realizations is a subsequent, optional step
based on the parameter distribution.
Computational cost of this step is essentially negligible.
In other words,
computational cost of the inverse procedure is not tied to the
number of field realizations that are ultimately needed for prediction
tasks.

Anchored inversion is not centered at optimization.
Optimization-based approaches typically places much emphasis on
achieving certain level of model reproduction of the
observed data; this is not the case with the proposed method.
This distinction has important conceptual implications.
It avoids the difficult tasks of determining relative weights for the
performance and regularization terms, as well as relative weights for data
components in the performance term.
It also avoids the difficult task of determining a stopping rule for the
optimization algorithm,
and eliminates ad hoc statistical interpretations for realizations
of the attribute field
that terminate with different (or identical) values of an objective function.
This, naturally, also saves
much effort in developing an optimization algorithm.

The proposed method is general with regard to nonlinearity of
the forward model. Indeed, it views the linear case as a solved problem,
and is mainly targeted at incorporating data from nonlinear forward
processes.
The method (as is presented here)
assumes the forward model is deterministic,
but otherwise arbitrary.

Following this introduction,
in Section~\ref{sec:formulation} we present the formulation
of the method.
This section introduces the concept of ``anchor parameters''
or ``anchors'', which provides flexible parameterization of the field,
especially for local features, in addition to conventional
``structural parameters''.
This section also describes a data classification scheme.
Based on this scheme,
all data are incorporated systematically,
as is described in Section~\ref{sec:inference}.
The model inference presented in Section~\ref{sec:inference}
treats model and data errors as an integral part.
In Section~\ref{sec:sampling} we present a sampling-based method
for deriving the distribution of the model parameter vector.
The method has promising efficiency features,
and is a general idea in its own right.
Section~\ref{sec:illustrations} presents a synthetic study of a
one-dimensional problem, demonstrating model formulation and parameter
inference.

In Section~\ref{sec:connections} we discuss connections of anchored
inversion to existing methods in stochastic hydrogeology.
One emphasis is the popular ``pilot point'' method, which has
similarities to anchored inversion in terms of parameterization.
Another emphasis is the decomposition of error into measurement error
and model error.
This decomposition is conceptually and practically important,
but appears to be lacking in the existing methods discussed.

The paper concludes in Section~\ref{sec:summary}
with a summary of contribution and future work.

As for notation,
bold symbols are used for vectors, and nonbold symbols for scalars.
Matrices are upper-case bold letters.
The scalar symbol $\x$ is used for a single spatial location,
even if the space is multi-dimensional;
the bold symbol $\vx$ stands for a vector of locations.
All vectors are column vectors.
The concatenation of two vectors $\vec{a}$ and $\vec{b}$ is
$(\vec{a}^T, \vec{b}^T)^T$,
where the superscript $\scriptstyle{T}$ denotes matrix transpose,
but for simplicity we write
$(\vec{a}, \vec{b})$.
We assume the field is discretized for modeling purposes;
as a result we will be able to use matrix notations instead of
integrals.
Where the context permits,
we use a slight notational abuse to distinguish functions by the
independent variable.
For example,
$p(\phi)$ and $p(\psi)$ are not the same function evaluated at two
values but are two different functions, of $\phi$ and $\psi$
respectively.

\section{Model formulation and data classification}
\label{sec:formulation}

Denote the spatial random process by $\Y(\x)$,
where $\Y \in \mathfrak{R}$ is a variable of interest
indexed by the spatial coordinate $\x$ in
1-, 2- or 3-D space.
We model the spatial variable $\Y$ as a Gaussian process.
In conventional geostatistical formulation,
this process may be described by a ``structural parameter'' vector,
$\vtheta$,
comprising two groups of parameters:
parameters that define the expected value (mean function) of $\Y$ at any
location, and parameters that describe the association (covariance)
of $\Y$'s at different locations.
In essence, this formulation assumes
\begin{equation}\label{eq:gauss-field}
\vY(\vx) \given \vtheta
\sim
\gauss\bigl(\vmu(\vx), \, \mQ_{\vY,\vY}\bigr)
,
\end{equation}
where
$\vmu$ is the mean function
and
$\mQ$ is the covariance matrix,
both being functions of the location vector $\vx$
and the structural parameters $\vtheta$.
The inverse methodology being proposed makes no assumption about the
structural parameter $\vtheta$;
the specific form that is used to create the illustrations in this study
is presented in Appendix~B.
We use $\vfx$ to denote the location vector of the model grid of the
whole field, then
$\fY = \vY(\vfx)$ is the field vector of $\Y$ values.

If we know the value of a linear function of the field, say,
\[
\vvartheta = \mH \vfy
,
\]
where $\mH$ is a \emph{known} matrix and $\vfy$ is the vector of the
$\Y$ values in the whole field,
then a basic property of the normal distribution states
(see Appendix~A) that, conditional on this information,
the process $\Y$ still has a normal distribution:
\begin{equation}\label{eq:formulation}
\vY(\vx) \given (\vtheta, \vvartheta)
\sim
\gauss\bigl(
    \vmu(\vx) + \mQ_{\vY,\mH\fY}\, \mQ^{-1}_{\mH\fY,\mH\fY}
        \bigl(\vvartheta - \mH\vmu(\vfx)\bigr),
    \,
    \mQ_{\vY,\vY} - \mQ_{\vY,\mH\fY} \mQ^{-1}_{\mH\fY,\mH\fY}
        \mQ_{\mH\fY,\vY}
    \bigr)
,
\end{equation}
where
\begin{equation}\label{eq:formulation-Qs}
\mQ_{\vY,\mH\fY}
= \mQ_{\vY,\fY} \mH^T,\quad
\mQ_{\mH\fY,\mH\fY}
= \mH \mQ_{\fY,\fY} \mH^T,\quad
\mQ_{\mH\fY,\vY}
= \mH \mQ_{\fY,\vY}
.
\end{equation}
The superscript $\scriptstyle{T}$ denotes matrix transpose,
and $\mQ_{\vY,\fY}$ (or $\mQ_{\fY,\vY}$)
denotes the covariance matrix between
$\vY(\vx)$ and $\fY$ (or between $\fY$ and $\vY(\vx)$).

If we do not have such information as a linear function of the field,
however, the handy property above prompts us to designate a (wisely
chosen) linear function of the field as something special:
\[
\vvartheta \overset{\text{def}}{=} \mH\fY
,
\]
and treat $\vvartheta$ as unknown parameters.
(The matrix $\mH$ above has no relation to the previous $\mH$.)
We call these parameters ``anchors''.
Hence our model for the spatial process $\Y$
consists of the parameter vector
$\vTheta = (\vtheta, \vvartheta)$
and the user-specified matrix $\mH$.
The anchor parameters introduce variations in the mean function and
spatial correlation structure, as in~(\ref{eq:formulation}),
that are beyond the expressing capability of the structural parameters.
One may say that the structural parameters capture global features,
whereas the anchor parameters capture ``local'' (or additional global)
features.
One may also say that the structural parameters describe the field
prior to knowledge of the anchor parameters.

Given data $\vz$ that relate to $\Y$ in some way,
our goal in this study is to derive the conditional distribution
of the model parameter vector, that is, $p(\vTheta \given \vz)$.
Once this distribution has been obtained,
the distribution of the field in this model formulation is
\begin{equation}\label{eq:formulation-conditional-field}
p\bigl(\vfy \given p(\vTheta \given \vz)\bigr)
= \int p(\vfy \given \vTheta)\, p(\vTheta \given \vz) \diff\vTheta
,
\end{equation}
where $p(\vfy \given \vTheta)$ is given in~(\ref{eq:formulation})
(where $\vx$ is replaced by $\vfx$ and $\vY$ replaced by $\fY$).
Note that the distribution~(\ref{eq:formulation-conditional-field})
is \emph{not}
$p\bigl(\vfy \given \vz\bigr)$, which would be
$\int p(\vfy \given \vTheta, \vz)\, p(\vTheta \given \vz)
    \diff\vTheta$.
However,
the main point of the proposed method is to make
$p\bigl(\vfy \given p(\vTheta \given \vz)\bigr)$
in effect approximate
$p(\vfy \given \vz)$
for the purpose of predicting a particular field function,
say $f(\vfy)$, in the sense that
$\int p(f \given \vTheta)\, p(\vTheta \given \vz) \diff\vTheta
\approx p(f \given \vz)$.
This approximation is achieved by the method as a whole and
in particular by effective choice of the anchor parameters,
rather than by any single step of mathematical manipulation.
The field $\vfy \given p(\vTheta \given \vz)$ is conditioned on the data
$\vz$ \emph{indirectly} via the model parameters $\vTheta$.
By giving up direct control over this data conditioning
(one example of direct control is insisting on a specified level
of data reproduction by simulated fields),
the proposed method gains in
rigorous parameter inference and tractable statistical analysis
of the parameters as well as of the generated field realizations.
These benefits will become clear in subsequent sections.
We mention one benefit here:
the distribution $p(\vfy \given \vTheta)$ is the normal
distribution~(\ref{eq:formulation}) and is easy to sample from.
In contrast,
the distribution $p(\vfy \given \vTheta, \vz)$ is unknown and
there is no easy way to sample from it.

Now a natural question arises:
where does $\mH$ come from?
Or, how do we define or choose anchor parameters?
To answer this question (only partially in this paper),
we examine the various situations in the data-anchor relation
and classify all on-site data into two categories based on this
relation,
as follows.

\textbf{Type A data}: data that evaluate a linear function of
the field $\fY$:
\begin{equation}\label{eq:type-A}
    \vza
    = \mH\ped{a}\vfy + \verra
    ,
\end{equation}
where $\verra$ are errors with distribution
$p_{\verra}(\verra)$.
This data type includes two situations:
data of point values of $\Y$ at certain locations,
and data of linear functions of $\Y$
at multiple locations or the entire field.
The values $\vza$ may be measured directly,
or may be obtained through intermediate models of arbitrary
complexity.
The errors $\verra$
combine those from measurements and from any
intermediate models,
such as regressions and empirical relations.
In the case where empirical relations give a nonlinear function of
$\Y$ at a single location,
we transform the information to a linear function of
$\Y$, with the error distribution changed accordingly.

\textbf{Type B data}: data that evaluate a nonlinear function of $\Y$
at multiple locations (\eg, in a spatial domain):
\begin{equation}\label{eq:type-B}
    \vzb = \fM(\tilde{\vy}) + \vec{\epsilon}\ped{b}
    ,
\end{equation}
where $\fM$ is a known nonlinear function
embodied in either analytical forms or numerical models.
In the context of ``inversion'',
$\fM$ is called the ``forward function'' or ``forward model''.
The errors $\verrb$, with distribution
$p_{\verrb}(\verrb)$, combine measurement errors
and errors in the implementation of the function $\fM$
(see Section~\ref{sec:errors}).

With type-A data, we always define corresponding anchors
\[
    \vvarthetaa \overset{\text{def}}{=} \mH\ped{a} \fY
.
\]
The values of these anchor parameters are directly provided
by the data $\vza$, possibly with uncertainty.
In order to capture the information in type-B data,
we define additional anchors
\[
    \vvarthetab \overset{\text{def}}{=} \mH\ped{b} \fY
,
\]
where the linear function (\ie matrix) $\mH\ped{b}$
is determined with or without reference to the forward function $\fM$.
We call $\vvarthetaa$ and $\vvarthetab$ anchors of type A and B,
respectively;
or, more descriptively, the former ``measured anchors''
and the latter, ``inverted anchors''.
If an anchor is defined as the value of $\Y$ at a specific location,
we may call it an ``anchor point''; otherwise an ``anchor function''.
Collectively, we write
$\vvartheta = (\vvarthetaa, \vvarthetab)$ and
$\mH = \begin{bmatrix}\mH\ped{a}\\ \mH\ped{b}\end{bmatrix}$.

The simplest way of defining type-B anchors is to
choose unsampled locations $\{\x_i\}$
and designate the variables $\{\Y(\x_i)\}$ as anchors.
Generally speaking, we should choose locations $\{x_i\}$ that significantly
increase the resolving power of the formulation~(\ref{eq:formulation})
for the field.
An interesting situation happens when the forward function
$\fM$ is close to a linear function of the field,
say $\vec{h}(\vfy)$.
In that situation,
we designate $\vec{h}(\fY)$ (with a known form) as anchor parameters.
By this designation, the data naturally are very informative
about these anchor parameters,
although they do not provide the parameter values directly.

Further detail regarding the strategy of defining type-B anchors is
beyond the scope of this article.
It is worth mentioning that inverted anchor points are analogous to the
``pilot points'' in \citet{deMarsily:1984:IIT}
(see Section~\ref{sec:PP} for more discussion).
In the context of the pilot point method,
\citet{RamaRao:1995:PPM1} and \citet{Gomez-Hernanez:1997:SST1} discuss
strategies for choosing locations as pilot points.

In the groundwater example in Section~\ref{sec:intro},
direct measurements of local-scale hydraulic conductivity
or transmissivity (always with much uncertainty) are type-A data.
Type-A data also include
covariates such as grain-size distribution and
core-support geophysical properties.
Examples of type-B data are
head measurements in well pumping tests and
concentration measurements in tracer transport experiments.

In the geophysical example in Section~\ref{sec:intro},
type-A data are mechanic properties of the elastic medium,
usually only available at the ground surface.
Type-B data are recorded pressure and particle velocity at the ground
surface.

In the atmospheric example in Section~\ref{sec:intro},
type-A data are direct monitoring of CO$_2$ sources on the ground
surface or covariates that provide estimates of CO$_2$ source.
Type-B data are CO$_2$ concentration measured mainly on the ground
surface.

An important issue in type-A data is ``change of support (or scale, or
resolution)''.
For example,
\citet{Merlin:2005:CMM} develop a disaggregation method to use
40\unit{km} resolution satellite data, along with 1\unit{km} auxiliary data,
in fine scale modeling of surface soil moisture.
\citet{Bindlish:2002:SVR} describe a method to combine
two remote sensing datasets of 30\unit{m} and 200\unit{m} resolutions, respectively,
in modeling surface soil moisture.
\citet{Fuentes:2005:MES} develop methods to combine station monitoring and
regional-scale model output data for air pollution;
beneath the point-scale measurements and regional-scale predictions,
both subject to error, is a random field defined in continuous space.

\section{Inference of the model parameters}
\label{sec:inference}

In this section we derive the parameter distribution
conditional on all data, that is,
$p(\vTheta \given \vz)$.
The derivation contains the following three steps.

First, we assign a prior for the parameter vector in the form
\begin{equation}
p(\vTheta)
= p(\vtheta)\, p(\vvartheta \given \vtheta)
.
\end{equation}
Because of the Gaussian assumption~(\ref{eq:gauss-field}),
the conditional specification
$p(\vvartheta \given \vtheta)$ is a normal distribution (given in Appendix~A).
Only $p(\vtheta)$ requires to be specified by the user.
The specification of this term is transparent to the anchored inversion
methodology; the particular form we have used is given in Appendix~B.

Second, we derive the conditional distribution of the parameter vector
given type-A data, that is,
$ p(\vTheta \given \vza, p_{\verra}) $.

Third, we update the preceding conditional distribution to be
conditioned on type-B data, that is, derive
$ p(\vTheta \given \vza, p_{\verra},
    \vzb, p_{\verrb})$.

While the three steps are a coherent sequence,
the last step is the main contribution of this study
(besides the anchor formulation).

\subsection{Using type-A data}
\label{sec:use-type-A-data}

Noticing that the type-A data $\vza$ and $p_{\verra}$
simply stipulate the likelihood of the type-A anchors $\vvarthetaa$
independently of other parameters,
we have
\begin{equation}\label{eq:Theta-given-za}
p(\vTheta \given \vza, p_{\verra})
\propto p(\vTheta)\, p_{\verra}(\vza - \vvarthetaa)
.
\end{equation}

In the special case where the data $\vza$ are free of error, that is,
$p_{\verra} = \delta(\vec{0})$,
the conditional distribution~(\ref{eq:Theta-given-za}) is equivalent
to
$\vvarthetaa = \vza$ and
\begin{equation}\label{eq:Theta-given-za-noerr}
p(\vtheta, \vvarthetab \given \vvarthetaa)
\propto
    p(\vtheta)\,
    p(\vvarthetaa \given \vtheta)\,
    p(\vvarthetab \given \vtheta, \vvarthetaa)
,
\end{equation}
in which $\vvarthetaa = \vza$.
Both
$p(\vvarthetaa \given \vtheta)$ and
$p(\vvarthetab \given \vtheta, \vvarthetaa)$
are normal distributions (see Appendix~A).

Assume we have a way to sample the parameter
distribution~(\ref{eq:Theta-given-za-noerr}) (see Appendix~B),
then sampling the distribution~(\ref{eq:Theta-given-za}) is simple:
first sample a random error $\verra^*$ from $p_{\verra}$,
then set $\vvarthetaa = \vza - \verra^*$
and sample $(\vtheta, \vvarthetab)$
from~(\ref{eq:Theta-given-za-noerr}).

\subsection{Using type-B data in addition to type-A data}
\label{sec:use-type-B-data}

Via the Bayes theorem we have
\begin{equation}\label{eq:Theta-given-zab}
\begin{split}
p(\vTheta \given \vza, p_{\verra}, \vzb, p_{\verrb})
&\propto
    p(\vTheta \given \vza, p_{\verra})
    \times
    p(\vzb, p_{\verrb} \given \vTheta, \vza, p_{\verra})
\\
&\propto
    p(\vTheta) \,
    p_{\verra}(\vza - \vvarthetaa)
    \times
    p(\vzb, p_{\verrb} \given \vTheta)
\\
&=
    p(\vTheta) \,
    p_{\verra}(\vza - \vvarthetaa)
    \times
    \int p_{\fM} \bigl(\vzb - \verrb \given \vTheta\bigr)\,
        p_{\verrb}(\verrb) \diff \verrb
.
\end{split}
\end{equation}
(The equality
$p(\vzb, p_{\verrb} \given \vTheta, \vza, p_{\verra})
= p(\vzb, p_{\verrb} \given \vTheta)$ holds because
$\vTheta$ contains a specific value of $\vvarthetaa$,
which overrides $\vza$ and $p_{\verra}$.)
The likelihood term
$p_{\fM}\bigl(\vzb - \verrb \given \vTheta)$
connects the type-B data $\vzb$ to the model parameter $\vTheta$.
This likelihood is the probability density function
of the random variable $\fM(\vfy) \given \vTheta$,
evaluated at $\vzb - \verrb$.
The distribution of $\fM(\vfy) \given \vTheta$
is determined by $\vTheta$ and $\fM$.
In general we do not have an analytical form for this distribution;
neither is there a basis for assuming any convenient form for it.
At this point,
we make no assumption about the nature of the forward function $\fM$
other than that it is known and deterministic.
It can be, for example,
a simple combination of multiple field functions
that generate multiple datasets of incomparable nature
(such as travel time and chemical concentration).
To keep the methodology general,
we do not expect to be able to calculate the likelihood
$p_{\fM}\bigl(\vzb - \verrb \given \vTheta)$ analytically.

In principle, this likelihood can be estimated numerically and
nonparametrically.
The reason is that we can simulate the random variable
$\fM(\vfy) \given \vTheta$ by
sampling $\vfy^*$ from $p(\vfy \given \vTheta)$
and then evaluating $\fM(\vfy^*)$.
This simulation provides a random sample of
$\fM(\vfy) \given \vTheta$,
therefore the density function of this variable can be estimated
nonparametrically.
In particular, the density at $\vzb - \verrb$
is the likelihood $p_{\fM}\bigl(\vzb - \verrb \given \vTheta)$
in~(\ref{eq:Theta-given-zab}).

Nonparametric density estimation is a large topic in
statistical literature \citep{Scott:1992:MDE, Wand:1995:KS}.
The density in the targeted applications of this study is
usually high-dimensional, because the dimensionality is the length
of the data vector $\vzb$.
Estimation of high-dimensional densities is a notoriously difficult
task, due to the so-called ``curse of dimensionality''
\citep[see][sec.~1.5 and chap.~7]{Scott:1992:MDE}.
In Section~\ref{sec:sampling},
we propose a strategy that avoids evaluating the likelihood
$p(\vzb, p_{\verrb} \given \vTheta)$ altogether.

If the data $\vzb$ are free of error, then the integral
in~(\ref{eq:Theta-given-zab}) reduces to
$p_{\fM}\bigl(\vzb \given \vTheta\bigr)$.

\subsection{Using type-B data without type-A data}

If all data are of type B, then
all anchors are inverted ones.
In this case, the posterior parameter distribution is
\[
\begin{split}
p(\vtheta, \vvarthetab \given \vzb, p_{\verrb})
&\propto
  p(\vtheta, \vvarthetab)
  \times
  p(\vzb, p_{\verrb} \given \vtheta, \vvarthetab)
\\
&\propto
  p(\vtheta)\,
  p(\vvarthetab \given \vtheta)
  \times
  \int p_{\fM}\bigl(\vzb - \verrb \given \vtheta, \vvarthetab\bigr)
    \, p_{\verrb}(\verrb)
    \diff \verrb
.
\end{split}
\]
This proceeds as explained in Section~\ref{sec:use-type-B-data}.
A notable difficulty in this situation is the specification of a prior
for the structural parameters $\vtheta$.
In the case where type-A data are available,
the likelihood term
$p(\vvarthetaa = \vza \given \vtheta)$ provides much-needed regulation on
the structural parameters such that one can afford to use a
relatively naive prior for $\vtheta$.
One does not have this luxury in the absence of type-A data.
To appreciate the complication in specifying a prior,
see \citet{Kass:1996:SPD}, and \citet{Scales:2001:PIU}.

\subsection{Measurement error and model error: why doing it this way}
\label{sec:errors}

In this subsection we answer two questions:
(1) Why do we use a reduced parameter vector ($\vTheta$),
instead of the state space vector ($\vfy$)?
(2) Why do not we know the likelihood function
in~(\ref{eq:Theta-given-zab}) or assume one?

Let
$\vfy$ be a realization of the spatial field $\fY$.
In the modeling study,
it is necessarily assumed that this vector is indeed a good representation
of what is needed for predicting the forward process, $\fM_0$,
which is modeled (or approximated) by an analytical function or
numerical code $\fM$.
The true outcome in the field that is faithfully described by the
model input $\vfy$ is $\fM_0(\vfy)$,
while our model predicts the outcome to be
$\fM(\vfy)$.
The difference,
\[
{\verrb}_1 \overset{\text{def}}{=} \fM(\vfy) - \fM_0(\vfy),
\]
arising from the difference between $\fM$ and $\fM_0$,
is the error in the forward model.

In real-world applications model error always exists.
There are numerous sources for this error, including
spatial and temporal discretization,
inaccurate boundary and initial conditions,
relevant physical factors that are omitted from consideration,
deficiencies in the mathematical description of the physical process,
and so on.
Taken as a random variable,
the model error vector ${\verrb}_1$
usually has inter-dependent components.
Furthermore,
the distribution of this error vector may be dependent
on the input field $\vfy$;
imagine a realistic scenario where,
as the input $\vfy$ deviates more severely from
the actual field being studied,
the model $\fM$ becomes a less satisfactory representation of the actual
process $\fM_0$, hence the model error tends to be larger.
Considering these factors,
we conclude that model errors are elusive, yet may not be ``small''.

A conceptually different issue is the discrepancy between the
measurement of a quantity and its true value, written as
\[
{\verrb}_2 \overset{\text{def}}{=} \vzb - \fM_0(\vfy),
\]
where $\vzb$ is the measurement.
This measurement error arises from
intrinsic inaccuracies in the measurement instrument and technique,
uncertainties due to human operations,
and other random factors.
It is almost always acceptable to assume independence between
components of a measurement error vector.
It is also reasonable to assume that the measurement error is
independent of model error and model parameters
(including model input $\vfy$).
It is often the case that one has a better idea about measurement error
than about model error.

Somewhat between model and measurement errors
is error due to the so-called incommensurability,
\ie the fact that what is computed by the model
and what is measured in the field are not exactly the same quantity.
A typical example is the case where
a property is measured and modeled with different support (or resolution,
or scale).
In this discussion we include this error in the measurement error
(${\verrb}_2$) for convenience.

It is now clear that the discrepancy between data and model output
is the combined contribution of model and measurement errors:
\[
\verrb
= \vzb - \fM(\vfy)
= \bigl(\vzb - \fM_0(\vfy)\bigr) -
  \bigl(\fM(\vfy) - \fM_0(\vfy)\bigr)
= {\verrb}_2 - {\verrb}_1
.
\]
This is the error term in~(\ref{eq:type-B}).
In derivation~(\ref{eq:Theta-given-zab}) we need
$p_{\verrb}$, the distribution of $\verrb$.
Unfortunately,
this distribution is unknown in most applications,
thanks in no small part to the model error ${\verrb}_1$,
which eludes reliable quantification and yet may be as large
as or larger than the measurement error ${\verrb}_2$.

If we took a state-space approach,
treating the whole filed $\fY$ as the model parameter vector,
then the likelihood of this parameter vector with respect to the data
$\vzb$ and $p_{\verrb}$ would be
\[
p_{\verrb}\bigl(\vzb - \fM(\vfy) \given \vfy\bigr)
.
\]
As is explained above,
this likelihood function is unknown,
even if we assume it does not vary with the parameter vector $\vfy$.
The main complicating factor is the part of the difference
$\vzb - \fM(\vfy)$ that stems from model error.

The proposed method avoids this problem by taking a reduced parameter
vector, $\vTheta$. The likelihood is now
\[
\int p_{\fM}\bigl(\vzb - \verrb \given \vTheta\bigr)
    \, p_{\verrb}(\verrb) \diff \verrb
\]
as in~(\ref{eq:Theta-given-zab}).
This can be re-written as
\begin{equation}\label{eq:Theta-likelihood-zb}
p_z(\vzb \given \vTheta)
,
\end{equation}
where
$p_z(\cdot)$ is the density function of the random variable
$\fM(\vfy) + \verrb$.
Note the data $\vzb$ are the observation of this random variable,
according to~(\ref{eq:type-B}).
A crucial point is that $\fM(\vfy)$ is \emph{random} given a specific
$\vTheta$, because the (discretized) field $\vfy$ is random conditional
on $\vTheta$.
While
$\fM(\vfy) + \verrb$
is the sum of two random variables,
its distribution is approximately that of
$\fM(\vfy)$ if
we formulate the system such that variability of
$\fM(\vfy) \given \vTheta$ dominates that of $\verrb$.
In that case the likelihood~(\ref{eq:Theta-likelihood-zb})
is well defined by the behavior of $\fM(\vfy)$,
permitting us to ignore the elusive but smaller $\verrb$.

This formulation,
by reducing the model parameter dimensionality so as to
create more variability,
places statistical inference of the model parameter(s) on solid ground.
It eliminates ad hoc assessment of the model-data ``mismatch''
($\fM(\vfy) - \vzb$); meanwhile it is straightforward to accommodate
model and measurement errors if their distributions are both known
(which is unlikely).
The statistical relation between model parameters and observations
does not rely heavily on accurate quantification of model and
measurement errors.
The method works just fine, as it should,
in the limiting case where both model and
measurement errors vanish, which is the case in many synthetic studies.
This issue is further discussed in
Section~\ref{sec:common-likelihood-formulation}.

\section{Sampling the posterior distribution}
\label{sec:sampling}

It is generally impossible to get a closed form for the parameter
distribution $p(\vTheta \given \vz)$.
Inference on this posterior involves two high-dimensional problems.
The first is the dimensionality of the parameter vector $\vTheta$.
Usually there are below or close to 10 structural parameters including
trend coefficients, variance, scale and smoothness of the correlation
function, nugget, and anisotropy parameters.
There is no hard limit on the number of anchor parameters;
they can be several tens or more, depending on the amount of data
and other considerations.
(Here we only count the \emph{uncertain} anchor parameters.
Type-A anchors that are known exactly do not add to the complexity
of the problem.)
The second problem is the dimensionality of the likelihood (or density)
function $p(\vzb, p_{\verrb} \given \vTheta)$.
This dimensionality is the length of the type-B data vector $\vzb$,
which we expect to number in the tens.
We have mentioned that conceptually this likelihood can be estimated
after generating a large sample of the random variable
$\fM(\vfy) \given \vTheta$;
however, density estimation in such high dimensions is only next to
impossible \citep{Scott:1991:FMD},
noticing the high cost in running the forward model
$\fM$.

The high dimensionality of the model parameter vector
calls for a Monte Carlo approach,
whereby the posterior distribution is represented (\ie approximated) by
a discrete sample from it.
A popular method for this task is Markov chain Monte Carlo (MCMC).
In this study we do not use MCMC, mainly because the likelihood function
is unknown.
Nevertheless, a sampling-based method is the only option.

Since density estimation in more than, say, five dimensions
is practically hopeless,
a possible route is to use rough density estimation as a guide for the
sampling, which we have explored with mixed performance.
A related idea is a variant of MCMC
called ``approximate Bayesian computation (ABC)''
\citep[see][]{Marjoram:2003:MCM}.
We mention two challenges to the ABC method.
First, one has to deal with all the nontrivial technical issues in a
MCMC procedure, including burn-in, convergence, and so on.
Second, ABC uses a measure of the simulation-observation mismatch in
lieu of likelihood.
For multivariate data,
there are significant difficulties in how to define this measure.

The procedure we propose here is based on the following
(simple in hindsight) observations.
(1) From $p(\vTheta \given \vza)$ and
$p\bigl(\fM(\vfy) \given \vTheta\bigr)$ we have a joint distribution of
$\bigl(\vTheta, \fM(\vfy)\bigr) \given \vza$;
the target distribution $p(\vTheta \given \vza, \vzb)$
is a conditional of this joint distribution
at $\fM(\vfy) = \vzb$.
(2) If $p\bigl(\vTheta, \fM(\vfy)\bigr)$ is normal,
then the conditional is another normal distribution, known analytically.
(3) A normal mixture is very flexible and could approximate very
complicated distributions, such as $p\bigl(\vTheta, \fM(\vfy)\bigr)$
\citep{Marron:1992:EMI, West:1993:APD, Givens:1996:LAI,
McLachlan:2000:FMM}.

Let
$\{(\vTheta_i, w_i)\}_{i=1}^n$ be a sample of
$\vTheta \given \vza$ with weights $w_i$.
With the field parameterization and prior specification presented in
Appendix~B, it is straightforward to get a \emph{random} sample of
$\vTheta \given \vza$, hence the weights are all $1/n$.
In general,
we allow the sample to be obtained by any sophisticated algorithm with
proper weights; this is practical because it does not require evaluating
the expensive forward function $\fM$.
Subsequently,
randomly draw
$\vfy_i$ from $p(\vfy \given \vTheta_i)$ and evaluate
$\fM(\vfy_i)$, for $i = 1, \dotsc, n$.
We thus obtain a sample of the random vector
$\bigl(\vTheta, \fM(\vfy)\bigr) \big| \vza$,
denoted by $\vec\xi$ below, with weights $\{w_i\}$.
The distribution of $\vec{\xi}$ can be approximated by a normal mixture
in the usual setting of kernel density estimation
\citep[see introduction by][and references therein]{Sheather:2004:DE}:
\begin{equation}\label{eq:joint-normal-mixture}
p(\vec{\xi} \given \vza)
= \sum_{i=1}^n
    w_i\, \phi\bigl(\vec{\xi}; \vec{\xi}_i, h_i \mSigma_i\bigr)
,
\end{equation}
where $\phi(\cdot; \vmu, \mSigma)$ is the normal density function
with mean $\vmu$ and covariance matrix $\mSigma$,
and $h_i$ is a scaling factor (``bandwidth'', taken to be 1 here).
This approximation places a Gaussian kernel at every sample point
$\vec{\xi}_i$, with a covariance matrix $\mSigma_i$ taken to be the
sample covariance of, say, the 500 nearest neighbors
of $\vec{\xi}_i$ in the sample,
in the Mahalanobis sense,

Given the normal mixture
distribution~(\ref{eq:joint-normal-mixture}) for
$\vec{\xi}$,
and $\vzb$ as the observed value of $\fM(\vfy)$,
the conditional distribution of $\vTheta$ is another normal
mixture.
Take the partition
$
\mSigma_i
= \begin{bmatrix}
    \mSigma_{11,i} & \mSigma_{12,i}\\
    \mSigma_{21,i} & \mSigma_{22,i}
  \end{bmatrix}
$,
then (see \citet{West:1993:APD} and Appendix~A)
\begin{equation}\label{eq:conditional-normal-mixture}
p(\vTheta \given \vza, \vzb)
= \sum_{i=1}^n v_i\, \phi\bigl(\vTheta; \vec{m}_i, \mat{V}_i\bigr)
,
\end{equation}
where
$v_i \propto w_i\, \phi\bigl(\vzb; \fM(\vfy_i), \mSigma_{22,i}\bigr)$,
normalized to sum to unit,
and
\[
\vec{m}_i
= \vTheta_i +
  \mSigma_{12,i} \mSigma_{22,i}^{-1} \bigl(\vzb - \fM(\vfy_i)\bigr)
,\quad
\mat{V}_i
= \mSigma_{11,i} -
  \mSigma_{12,i} \mSigma_{22,i}^{-1} \mSigma_{21,i}
.
\]

Uncertainty in the data $\vza$ may be accommodated while sampling $\vTheta$,
as described in Section~\ref{sec:use-type-A-data}.
Uncertainty in $\vzb$ may be accommodated by perturbing
the sampled $\fM(\vfy)$ according to a specification of the error
distribution, $p_{\verrb}$.

Because the posterior distribution of the model parameters is a normal
mixture, various statistics of the posterior can be calculated
analytically, without resorting to sampling the posterior.

To make the normal approximation defensible,
all components of the variable $\vec\xi$ should be defined on
$(-\infty, \infty)$.
In other words,
the procedure above does not derive the posterior distribution of the
parameter vector $\vTheta$, but of a transform of it.
Among the most useful transformations are the
log transformation for a positive variable
and the logit transformation for a doubly bounded variable,
$f(x) = \log\frac{x - \text{lower}}{\text{upper} - x}$.
For convenience of discussion,
transformations to the forward process outcome can be made part of
the forward function $\fM$.
The anchor parameters do not need a transformation since they are
normal variables by assumption or by definition.

Usually, evaluating the forward function $\fM$
(often requiring running a numerical model) is by far the
most expensive operation in the inversion.
In the procedure above,
the forward function is evaluated exactly once for each value ever
examined for the model parameter vector;
this happens when we evaluate $\fM(\vfy_i)$ for the field
$\vfy_i$ drawn from $p(\vfy \given \vTheta_i)$.
The conditioning operation in~(\ref{eq:conditional-normal-mixture})
forcefully brings information in the observation $\vzb$
into the statistical structure
of the system that has been revealed through sampling and simulation.
These features contribute to the efficiency of the procedure.

A challenge in this procedure is that the weights $\{v_i\}$
tend to be very skewed, that is, in a mixture of thousands of
components,
possibly only a few have significant weights.
This is a consequence of the high dimensionality of $\vec\xi$, or $\vTheta$.

\section{Illustrations}
\label{sec:illustrations}

\begin{figure}
\centering
\includegraphics{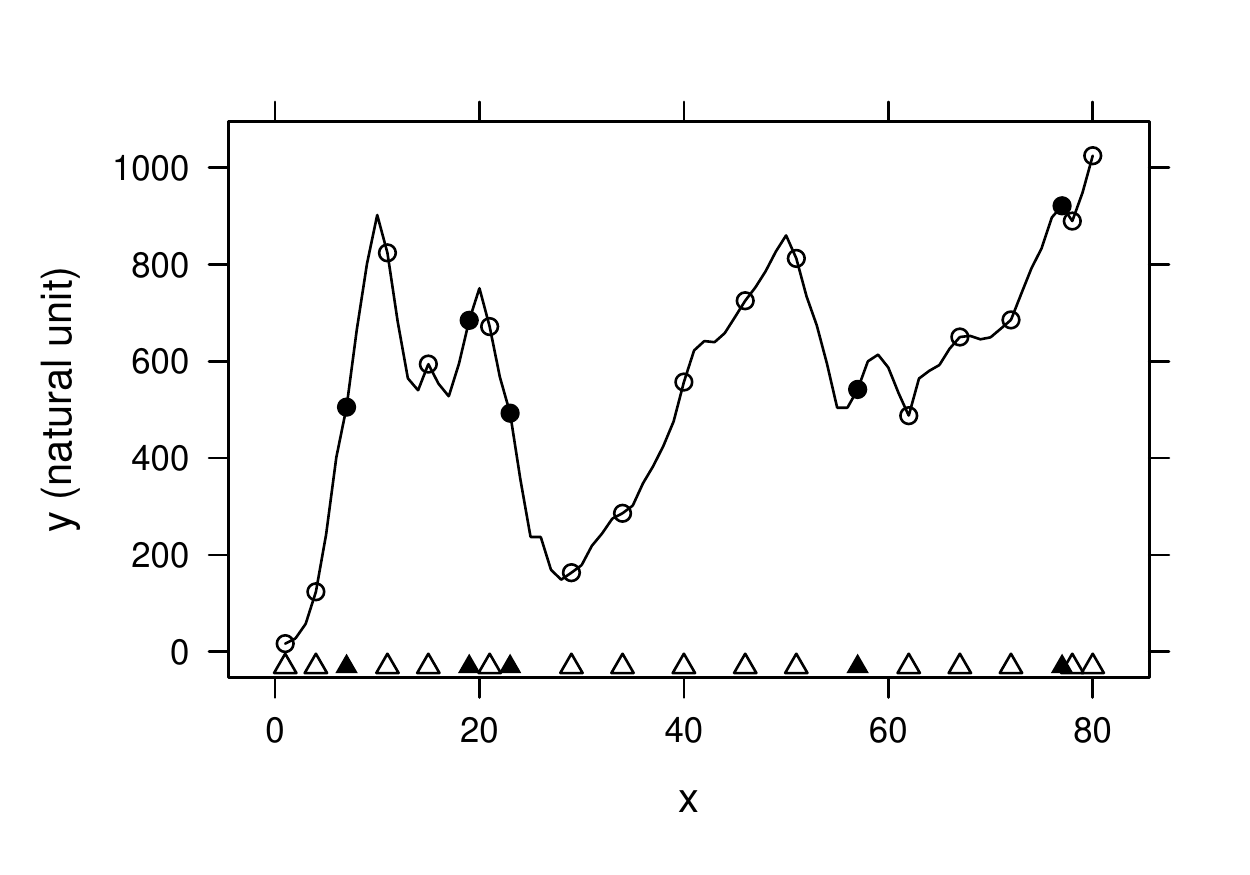}
\caption{True values of the 1-D field with 5 direct measurements
    (type-A data, filled symbols) and 15 inverted anchor points
    (unfilled symbols).}
\label{fig:true-field}
\end{figure}

We use a one-dimensional synthetic problem to illustrate the method.
The true 1-D field is shown in Figure~\ref{fig:true-field}.
This is actually the elevation data along some transect in the mountains
to the east of the city of Berkeley, California, USA.
The discretized profile is of length 80; the elevation is in meters.
We shall work in this discretized field of 80 locations,
and treat the elevation as representing some positive-by-definition
physical property ($\Y$ in its natural unit) that we need to infer.
Also marked out in Figure~\ref{fig:true-field} are 5 locations where we
have direct measurement of $\Y$ (\ie type-A data),
and 15 locations chosen as ``inverted anchors''.
We did not define anchor functions for this example:
all anchor parameters are anchor points.

\begin{figure}
\centering
\includegraphics{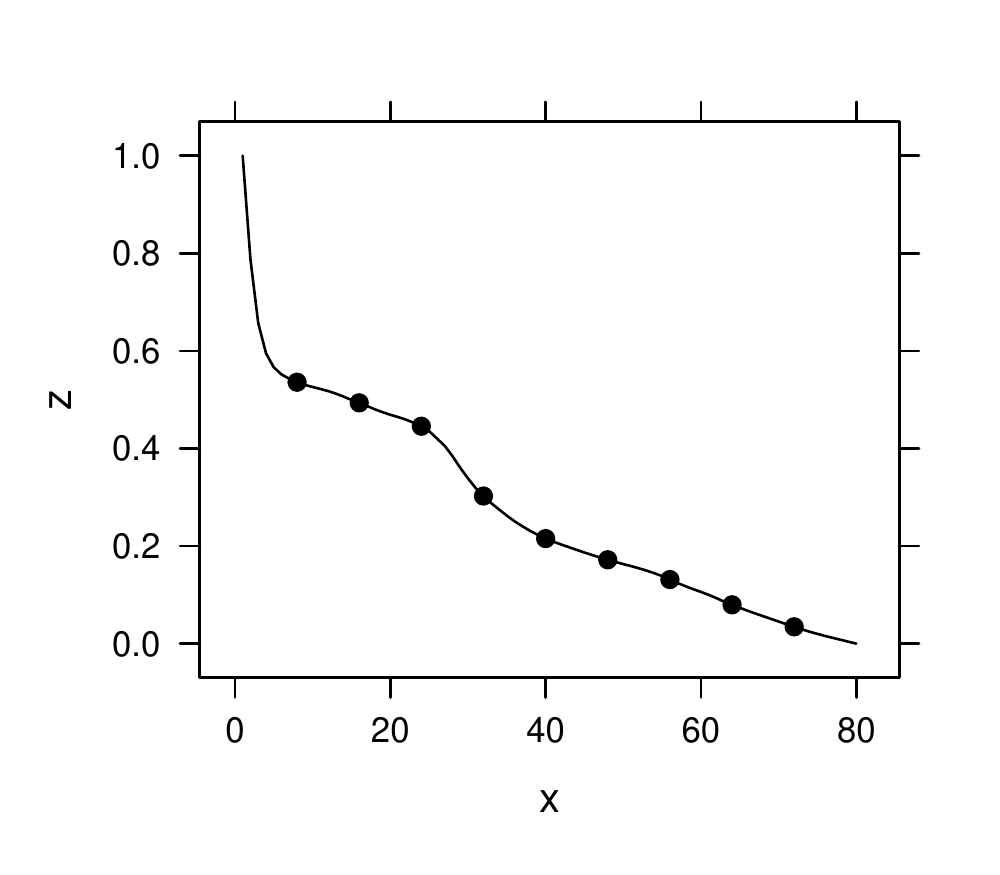}
\caption{The first synthetic forward process and 9 measurements of it.}
\label{fig:dataB-set1}
\end{figure}

\begin{figure}
\centering
\includegraphics{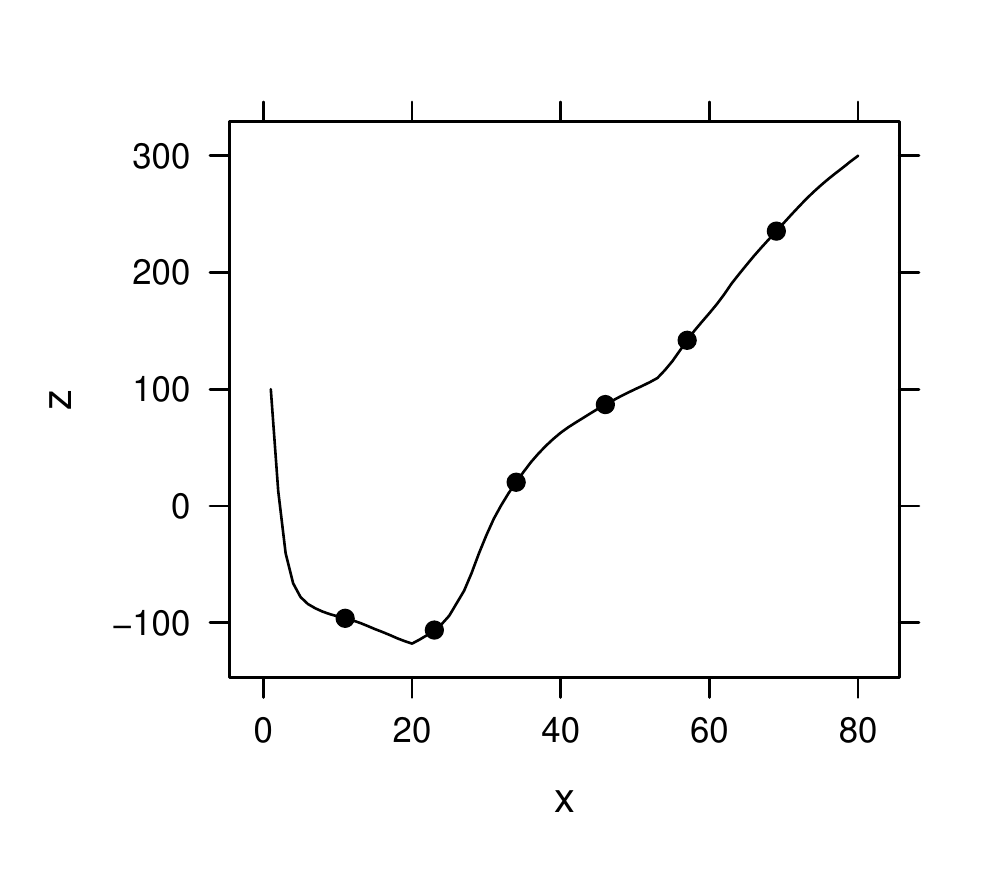}
\caption{The second synthetic forward process and 6 measurements of it.}
\label{fig:dataB-set2}
\end{figure}

We constructed two synthetic ``forward'' processes,
both described by the following differential equation:
\begin{equation}\label{eq:forward-synthetic}
\frac{\diff}{\diff x} \left(
    \y \frac{\diff z}{\diff x} \right) = s
,
\end{equation}
where $s$ is a source term,
$z$ is the forward process outcome controlled by
the field $\Y$ and the source $s$ as well as boundary conditions.
All of $y$, $z$, and $s$ are functions of location $x$.
The first forward process has $s = 0$ everywhere, $z(1) = 1$, and $z(80) = 0$.
The actual $z$ resulted from this process in the true field
is shown in Figure~\ref{fig:dataB-set1}, along with 9 measurements.
The second forward process has $s(20) = 4000$, $s(53) = 2000$,
$s = 0$ elsewhere, $z(1) = 100$, and $z(80) = 300$.
The actual $z$ along with 6 measurements are shown in
Figure~\ref{fig:dataB-set2}.
Note that we have intentionally made the two processes have
very different magnitudes.
The forward model $\fM$ in this setting
simply solves the differential equation separately with the two sets of
conditions, and combines the results to output a 15-vector.
The 15 measurements are type-B data $\vzb$,
whereas the 5 direct measurements are type-A data $\vza$.
All data were assumed to be error free.

We treated the unknown field as a Gaussian process,
assuming
a globally constant mean $\beta$,
an exponential covariance with scale parameter $\varphi$,
variance $\eta^2$,
and zero nugget.
Therefore the model parameter vector $\vTheta$
includes 3 structural parameters and 15 unknown type-B anchor parameters.
(Another 5 type-A anchor parameters are known.)
The posterior of $\vTheta$ was approximated by a normal mixture
of four sizes (the sample size $n$ in Section~\ref{sec:sampling}):
5000, 10000, 20000, and 40000,
referred to as scenarios AB1, AB2, AB3, and AB4, respectively.
Recall that the computational cost is proportional to the sample size.
The posterior distribution of the structural parameters was also
inferenced using type-A data only (see Appendix~B).
This posterior distribution was used along with the type-A data to
conduct simulations, referred to as scenario~A,
providing a reference for the AB scenarios.

While using the sampling algorithm described in Section~\ref{sec:sampling},
we transformed relevant variables as follows.
Scale parameter $\varphi$: logit transformation on $(5, 80)$,
\ie $\log \frac{\varphi - 5}{80 - \varphi}$.
Variance parameter $\eta^2$: log transformation.
Mean $\beta$: no transformation.
Forward model outcome: no transformation.
The inference of posterior described in
Section~\ref{sec:sampling} is really for these \emph{transformed} parameters.
In addition,
we took $\Y$ to be the logit transformation on $(1.7, 10249)$
of the elevation data.
(The value range of the elevation data is $[17, 1024.9]$.)
For convenience we say $Y$ is in ``model unit'' whereas
the positive field being modeled is in ``natural unit''.
Note the forward model~(\ref{eq:forward-synthetic}) takes
input field in its natural unit.

The prior specification for the structural parameters $\vtheta$
and the method for sampling the model parameters $\vTheta$
conditional on type-A data are described in Appendix~B.

\begin{figure}
\centering
\includegraphics[width=\textwidth]{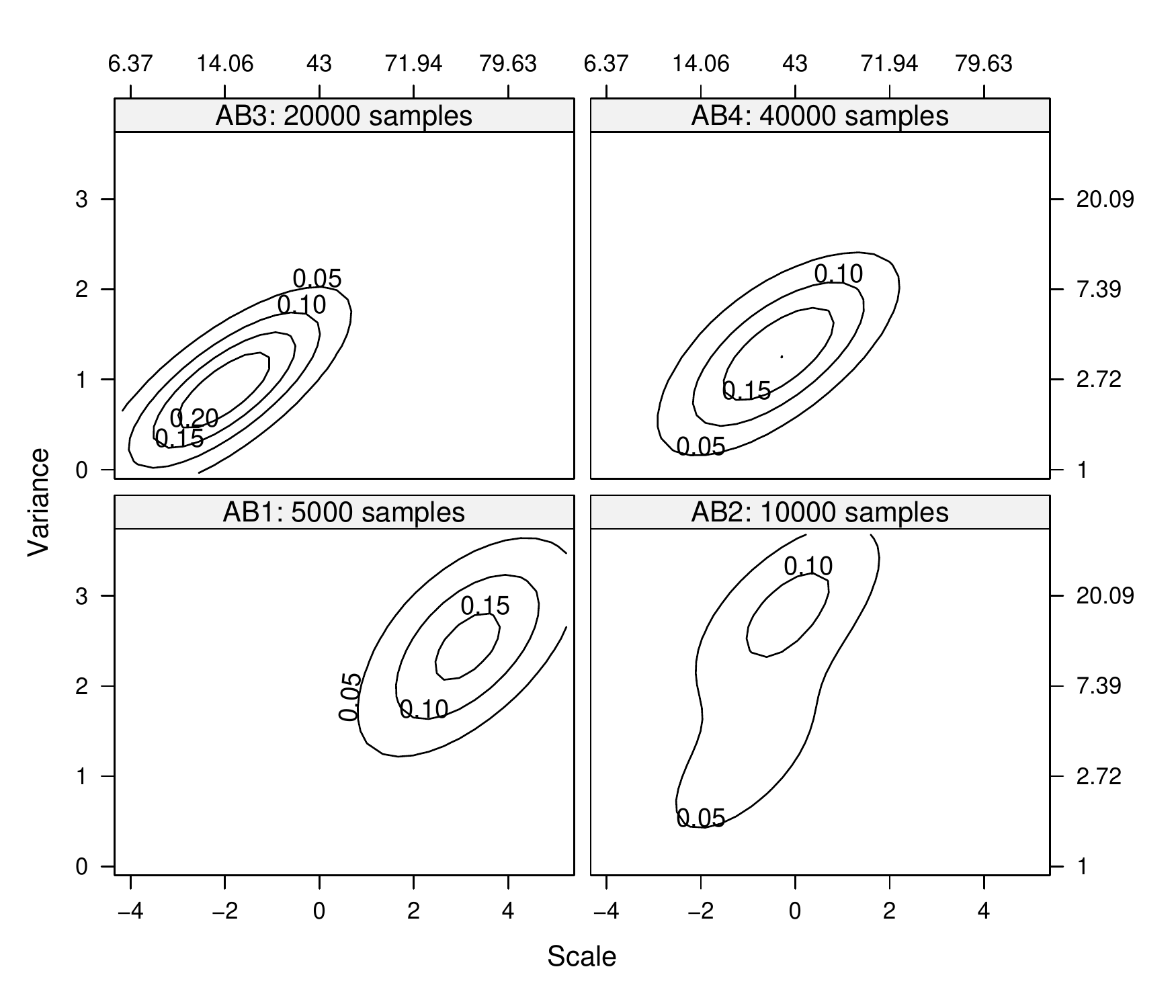}
\caption{Contour plot of the posterior joint density
of the transformed scale and variance parameters.
(Labels on the top and right are the back-transformed parameter
values.)}
\label{fig:scale-var}
\end{figure}

Figure~\ref{fig:scale} shows the posterior density of the scale
parameter,
calculated from the normal mixture approximation.
The posterior of scenario AB4 does not continue the trend of
decreasing scale in scenarios AB1--3,
suggesting some degree of convergence.
Because of the interaction between the scale and variance parameters
\citep[see][sec.~5.2.5]{Diggle:2007:MBG},
different value combinations of these two parameters
may have comparable model fitting performance.
This parameter interaction is clear in
Figure~\ref{fig:scale-var}, which shows
the joint posterior of this pair of parameters.
The variance parameter tends to be much smaller
in scenarios AB3--4 than AB1--2.

\begin{figure}
\centering
\includegraphics[width=\textwidth]{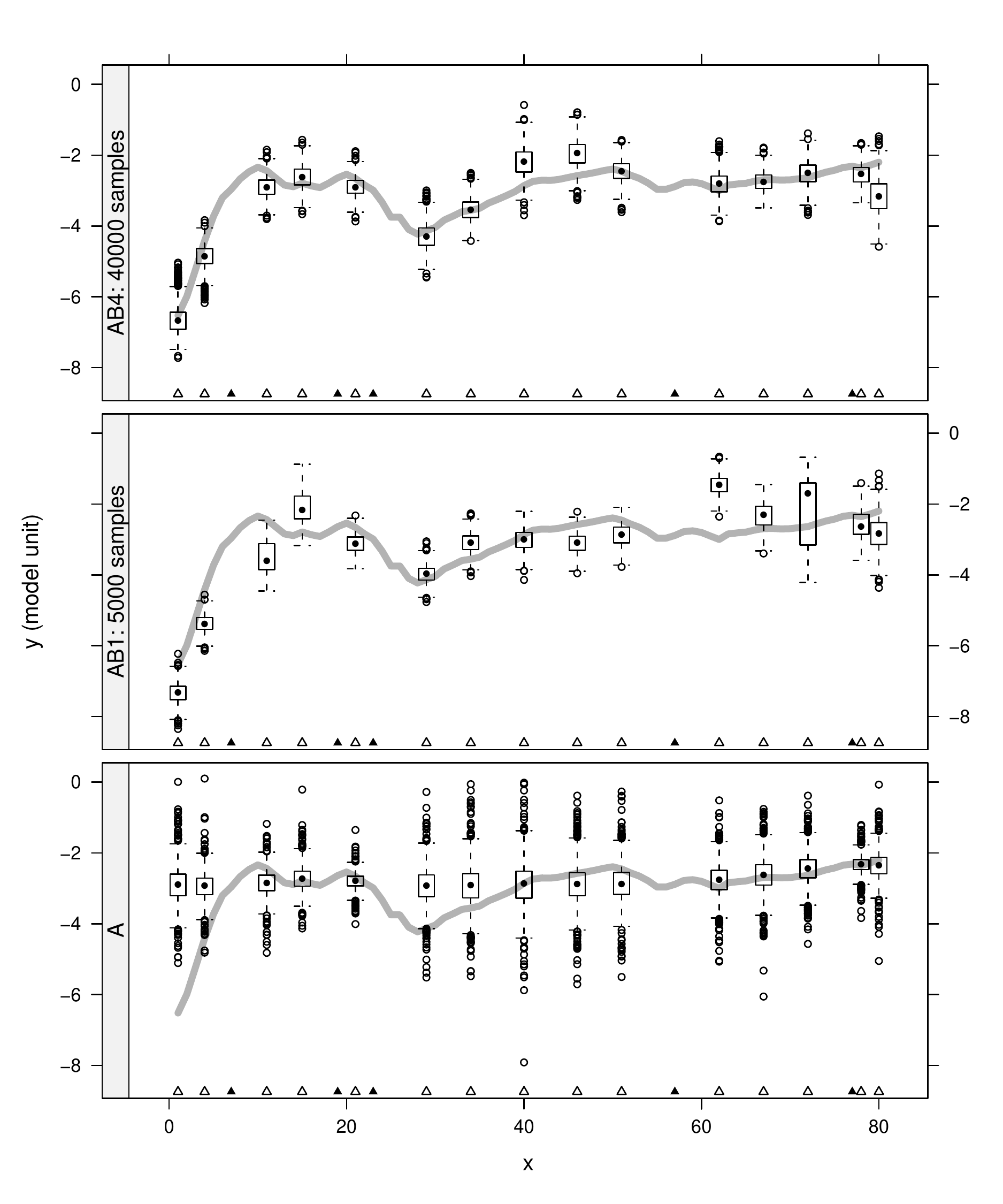}
\caption{Box plots showing the posterior marginal distributions of the
inverted anchors (in model unit), based on 1000 random draws
from the anchors' posterior distribution with respect to
type-A data (scenario A) and to both type-A and type-B data
(scenarios AB1 and AB4).
The thick gray curve is the actual field (in model unit).}
\label{fig:anchors}
\end{figure}

\begin{figure}
\centering
\includegraphics[width=\textwidth]{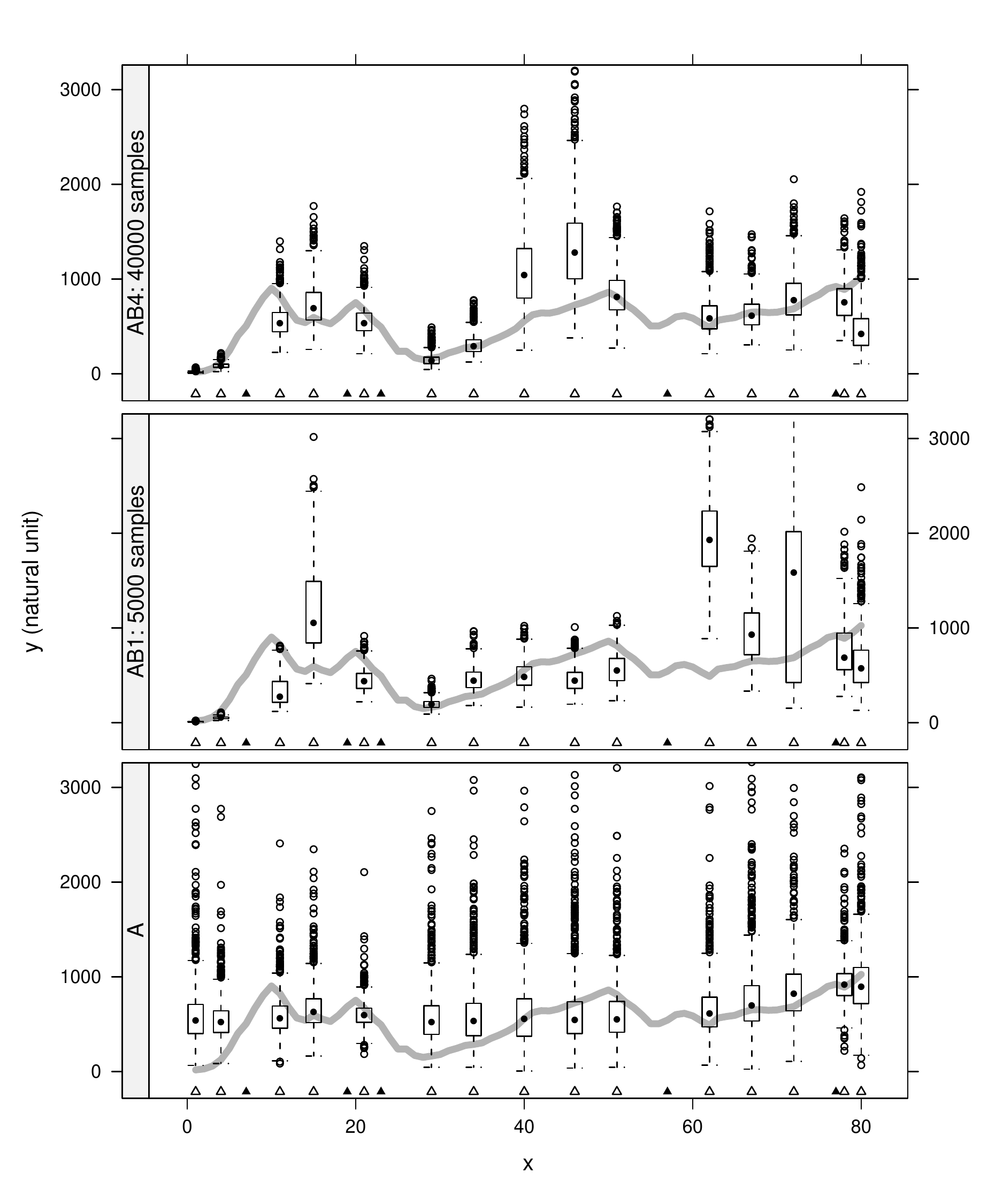}
\caption{Same as Figure~\ref{fig:anchors} but in natural units.}
\label{fig:anchors-natural}
\end{figure}

\begin{figure}
\centering
\includegraphics[width=\textwidth]{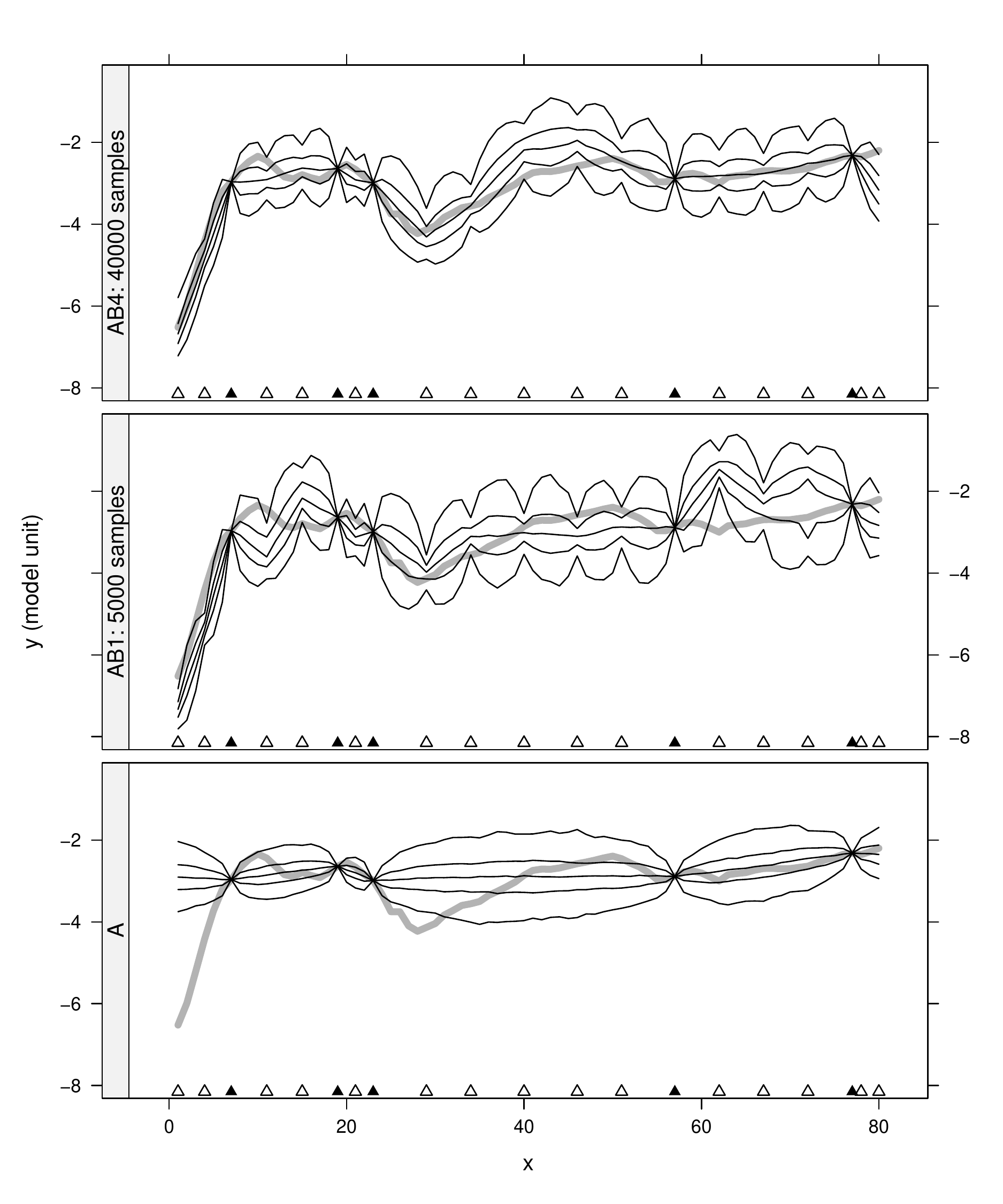}
\caption{The 5th, 25th, 50th, 75th, and 95th percentiles
of 1000 realizations of the field (in model unit)
created based on the parameter posterior with respect to
type-A data (scenario A) and to both type-A and type-B data
(scenarios AB1 and AB4).
The thick gray curve is the actual field (in model unit).}
\label{fig:fieldquarts}
\end{figure}

\begin{figure}
\centering
\includegraphics[width=\textwidth]{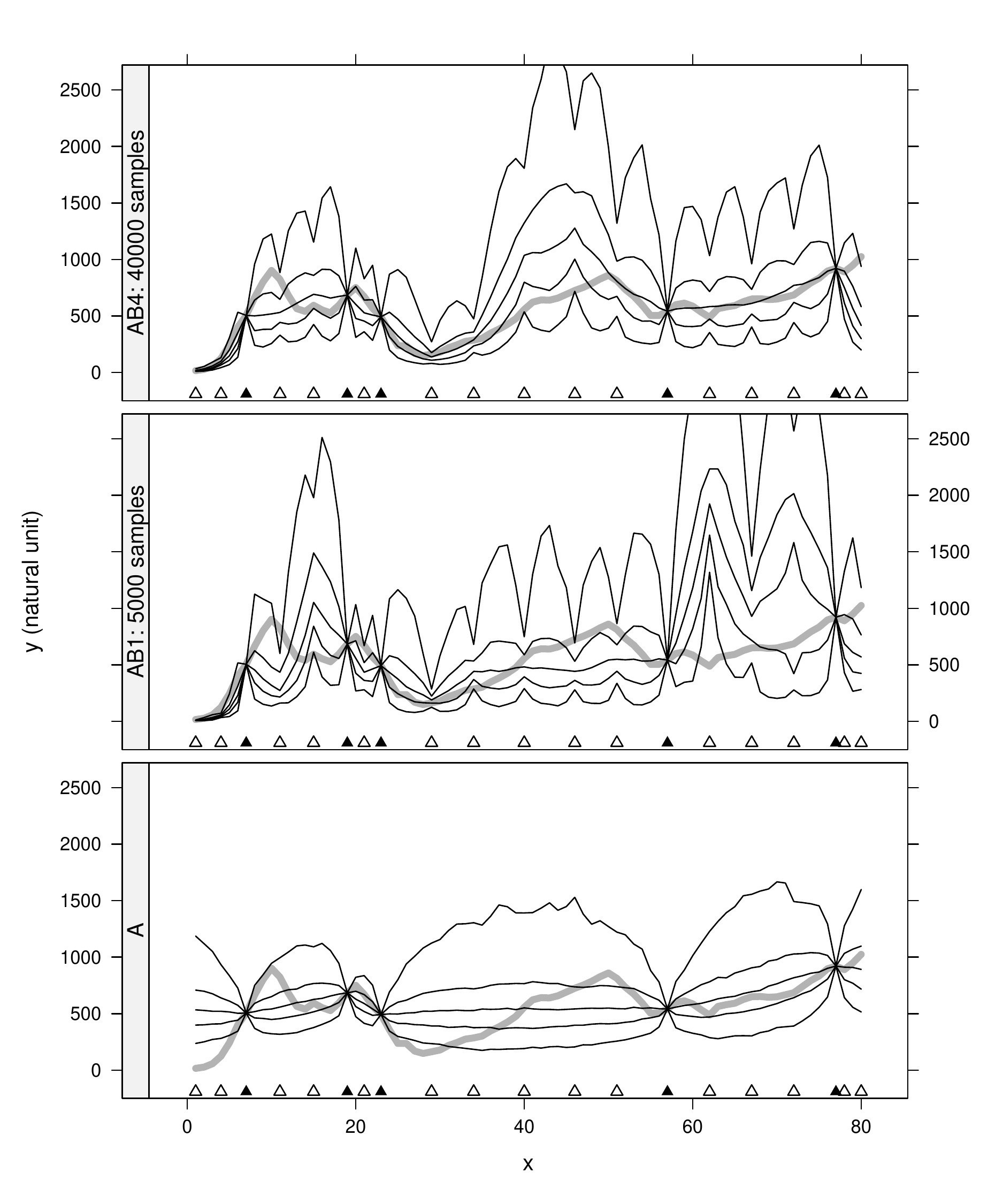}
\caption{Same as Figure~\ref{fig:fieldquarts} but in natural units.}
\label{fig:fieldquarts-natural}
\end{figure}

\begin{figure}
\centering
\includegraphics[width=\textwidth]{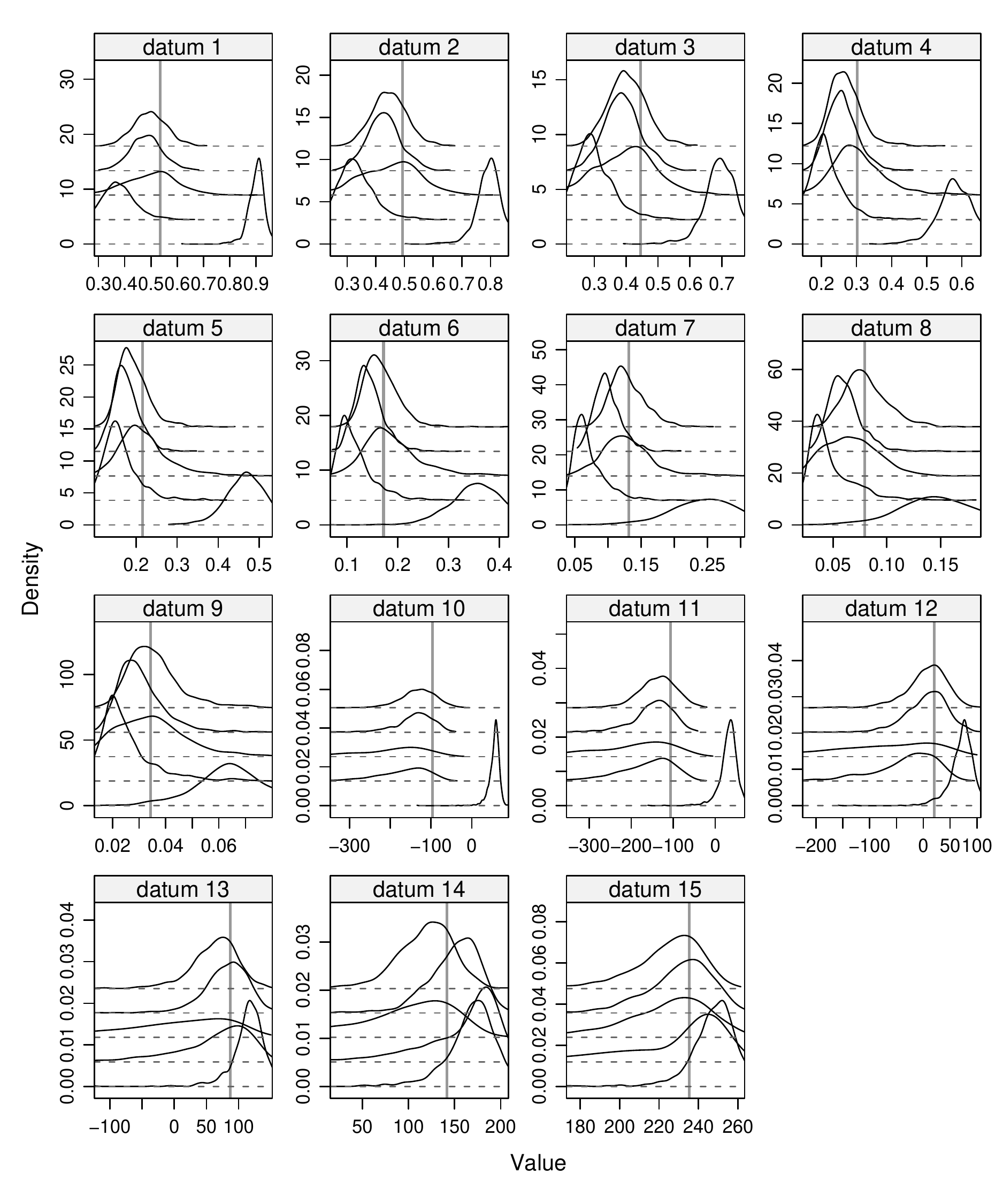}
\caption{Distributions of the forward model output
(\ie reproduction of the type-B data) in 1000 field realizations
created based on the parameter posterior with respect to
type-A data (bottom curve in each panel)
and to both type-A and type-B data (lifted curves in each panel;
with sample sizes 5000, 10000, 20000, 40000 from bottom to top).
The density values (on Y axes) should not be read literally except for
the bottom curve, due to the shifted stacking-up of multiple curves.
The vertical lines mark the true type-B data values.}
\label{fig:dataB-reproduce}
\end{figure}

For each AB scenario,
we randomly drew 1000 values for the model parameter vector from the
posterior distribution (\ie the normal mixture approximation),
and simulated one field realization based on each parameter value.
A prediction with the forward model was obtained with each simulated
realization.
Corresponding sampling and simulations were also done with scenario~A
for comparison.
These results are shown in
Figures~\ref{fig:anchors}--\ref{fig:dataB-reproduce}.

The distribution of the sample of inverted anchors are shown in
Figures~\ref{fig:anchors}--\ref{fig:anchors-natural},
with comparisons to the true anchor values and to scenario~A.
The AB scenarios are a clear improvement over scenario~A.
The scenario AB4 has reduced bias and variance compared to AB1.
Comparison of Figures \ref{fig:anchors} and~\ref{fig:anchors-natural}
reveals two problems.
First,
the deviation of the simulated anchor values from their true values
is amplified in the natural unit.
Second, in natural unit,
the posterior distribution of anchor values has a long tail upward.
If we were confident enough to impose a tighter range on the modeled
field values (we allowed it to be up to 10 times the maximum value of
the true field), both problems would be alleviated.
Figures \ref{fig:fieldquarts}--\ref{fig:fieldquarts-natural} show
ranges of the simulated fields and convey similar messages.

It is informative to examine how the simulated fields reproduce the
type-B data.
According to Figure~\ref{fig:dataB-reproduce},
the reproduction in scenarios AB2--4 is much better
than in scenario~A from a statistical perspective.
Note that
the two groups of data components (from two forward processes)
have drastically different magnitudes and
this fact is transparent to the proposed inversion method.

Since the data are error-free,
the non-exactness of the reproduction is solely a consequence of
the randomness in the model parameters and field simulations.
The ``sharpness'' of reproduction has to do with how sensitive each
datum is to the model formulation.
Compared to optimization-based methods that impose a specified tolerance
on the data reproduction, the observation-simulation mismatch in the
proposed method is unbounded in theory.
This should not be viewed as a weakness of the method;
an analogous situation is the Gaussian assumption on measurement errors.
In Figures \ref{fig:anchors}--\ref{fig:fieldquarts-natural},
only two of the four AB scenarios are shown due to space restrictions.
We intentionally showed scenario~AB1,
whose reproduction of type-B data according to
Figure~\ref{fig:dataB-reproduce} is not consistently better than scenario~A.
However, the comparison of scenarios AB1 and A in
Figures \ref{fig:anchors}--\ref{fig:fieldquarts-natural} makes the point
that there is much to examine than the reproduction of type-B data.

\section{Comparisons with other methods}
\label{sec:connections}

In this section we draw comparisons with some
literature in stochastic hydrogeology.
Emphasis is on conceptual points;
computational alternatives such as MCMC and Kalman filters
are not discussed.

\subsection{The pilot point method}
\label{sec:PP}

The Pilot Point (PP) method
\citep{deMarsily:1984:IIT, RamaRao:1995:PPM1, Gomez-Hernanez:1997:SST1,
Doherty:2003:GWM, Kowalsky:2004:EFP}
is a (generalized) least-square estimator of the field $\vfy$ defined as the minimizer
of the following objective function:
\begin{equation}\label{eq:PP-objective}
J = \bigl(\fM(\vfy) - \vzb\bigr)^T
    \mW^{-1}
    \bigl(\fM(\vfy) - \vzb\bigr)
    +
    (\vfy - \vmu)^T
    \mQ^{-1}_{\fY,\fY}
    (\vfy - \vmu)
,
\end{equation}
where $\mW^{-1}$ is a weight matrix,
$\mQ_{\fY,\fY}$
is the covariance between $\fY$ according to a spatial covariance
function,
and
$\vmu$ is a known reference (or prior mean) value of the field $\vfy$.
Both $\mQ$ and $\vmu$ are determined by pre-specified structural
parameters $\vtheta$.
This is the common ``regularized data reproduction'' formulation
for inverse problems.
The optimization algorithm begins with an initial field $\vfy_0$
drawn from the normal distribution
$\mathcal{N}(\vmu, \mQ_{\fY,\fY})$ and modifies the field until
$J$ falls below a prescribed threshold.
Optimizing realizations reached from different initial fields
are usually considered ``simulations'',
which present some idea about the uncertainty in the estimator.

Because optimizing the field vector $\vfy$ directly would be mostly
infeasible, a number of locations in the field are chosen as
``pilot points'', and the $Y$ values at these locations are used
to guide the optimization algorithm.
In so doing, the field is actually
\begin{equation}\label{eq:PP-formulation}
\vfy =
\vfy_0 + \mQ_{\fY,\vY^*}
        \mQ_{\vY^*,\vY^*}^{-1}
        \bigl(\vy^* - \vfy_0(\vx^*)\bigr)
,
\end{equation}
where
$\vfy^* = (\vfy^*\ped{d}, \vfy^*\ped{p})$
is a vector of (fixed) direct measurements $\vfy^*\ped{d}$
and (tunable) pilot point values $\vfy^*\ped{p}$,
and $\vx^*$ is the location vector of direct measurements
and pilot points.
The algorithm thus reduces the dimension of optimization
from the size of the model grid to the number of pilot points
in the context of a specific initial field $\vfy_0$ and known structural
parameters $\vtheta$.
In exchange for this dimension reduction,
the procedure fills in the $Y$ values at locations
other than $\vx^*$
based on the assumed spatial correlation ($\mQ(\cdot \given \vtheta)$).

Some studies make the choice of pilot points an internal matter of each
optimization run
\citep{RamaRao:1995:PPM1},
\ie contingent upon each particular $\vfy_0$;
others use the same set of pilot points across all optimization runs
\citep{Gomez-Hernanez:1997:SST1}.
In the former case,
the pilot points are purely an assistant to the optimization algorithm.
In the latter,
the pilot points permit a higher-level interpretation as model
parameters.

In a variant to the objective function in~(\ref{eq:PP-objective}),
the regularization term may be written in terms of
the pilot point values instead of the entire field vector.
The regularization term may also be omitted, in which case
the regularization is implicit in the procedure.

We shall make three observations on the PP method.

(1) PP is a state-space approach, in which the model parameter
vector is the entire field $\vfy$, which is obtained by optimization,
while the structural parameters $\vtheta$ and pilot points
$(\vx^*\ped{p}, \vy^*\ped{p})$ are intermediate devices to help
the optimization procedure.
Due to the nature of optimization,
the generated fields are not random samples from a well defined target
distribution.
As a result,
if one views the ensemble of realizations created in PP as ``simulations'',
it is unclear what distribution is being simulated from.
This distribution is collectively
determined by all aspects of the optimization procedure.
Zhang [2009, ``An analysis of the pilot point method in
stochastic hydrogeology'', in preparation]
provides an analysis of this distribution.

(2) The method requires the structural parameters $\vtheta$ to be
pre-specified, and has no provisions for systematically
updating these parameters in light of the data $\vzb$.
This inability is because these parameters are used in creating $\vfy_0$
and in updating the field (following modifications to the pilot point values)
in the optimization algorithm according to~(\ref{eq:PP-formulation}).
This is despite the claim of \citet{Kowalsky:2004:EFP} that
they can, but do not, optimize over the structural parameters the same
way as they do the pilot point values.

(3) Creating each realization goes through the same optimization
procedure, therefore computational cost of PP is proportional to
the number of realizations requested.

The proposed method of anchored inversion is a clear contrast to the PP
in all these three aspects.
In anchored inversion,
the model parameters are $\vTheta$, which
contains conventional geostatistical structural parameters and
anchor parameters;
the field $\vfy$ is formulated as having a normal distribution
described by $\vTheta$.
The distribution of $\vTheta$ conditional on all data is derived.
Given the posterior distribution of $\vTheta$,
creating realizations of $\vfy$ requires negligible additional
computation.

In a PP procedure, the pilot points may well be generalized
to be linear functions of the field, $\mH\vfy$,
in exactly the same way as in anchored inversion.
In fact, we expect this generalization to enhance
the efficiency of the optimization if the linear functions are well
chosen.
The difference between anchored inversion and the PP method
is not in this part of the parameterization, but in how this
parameterization is used.

\subsection{Two methods that estimate the mean and covariance of
the field conditional on the data}

The ``quasi-linear'' method \citep{Kitanidis:1995:QLG, Nowak:2004:MLM}
essentially parameterizes the field $\fY$ as a normal distribution
$\mathcal{N}(\vmu, \mSigma)$ with
\begin{align*}
\vmu
&= \vec{\xi} + \mQ_{\fY,\mH\fY}\, \mQ^{-1}_{\mH\fY,\mH\fY}
        \bigl(\vzb - \fM(\vec{\xi})\bigr)
\\
\mSigma
&= \mQ_{\fY,\fY} - \mQ_{\fY,\mH\fY} \mQ^{-1}_{\mH\fY,\mH\fY}
    \mQ_{\mH\fY,\fY}
.
\end{align*}
This formulation has two parameter vectors:
the distributed parameters $\vec{\xi}$ and the structural parameters
$\vtheta$.
The $\mQ$'s are covariance matrices defined by $\vtheta$,
analogous to~(\ref{eq:formulation-Qs}).
The coefficient matrix $\mH$
is the sensitivity matrix of the forward
process $\fM$ with regard to the mean field $\vmu$:
\[
\mH(\vmu) = \frac{\diff \fM(\vfy)}{\diff \vfy} \bigg|_{\vfy = \vmu}
.
\]
This definition makes it necessary that
the mean field $\vmu$ (or the parameters $\vec{\xi}$ and $\vtheta$)
be estimated in an iterative procedure,
updating $\mH$ as the estimate of $\vmu$ changes.
The parameters $\vec{\xi}$ and $\vtheta$ (hence
$\vmu$ and $\mSigma$) are estimated by optimizing an objective function
that involves data reproduction, regularization,
and functions of the structural parameters.
The linearization
$\fM(\vmu + \Delta) \approx \fM(\vmu) + \mH(\vmu) \Delta$
is instrumental in the optimization procedure.

After the estimates of $\vmu$ and $\mSigma$
have been obtained,
\citet{Kitanidis:1995:QLG} proposes to generate realizations
through an optimization procedure with an objective function
similar to~(\ref{eq:PP-objective})
(in which $\mQ$ is replaced by $\mSigma$).
We suspect this extra optimization is unnecessary and would choose
to sample from the normal distribution
$\mathcal{N}(\vmu, \mSigma)$ directly.
In fact,
\citet{Kitanidis:1995:QLG} shows that the extra optimization makes little
difference compared to direct sampling.

\citet{Hernandez:2006:ISM} use the pilot point parameterization
to describe the field as follows:
\[
\vfy =
\mX\vbeta + \mQ_{\fY,\vY^*}
        \mQ_{\vY^*,\vY^*}^{-1}
        \bigl(\vy^* - \mX_{\vx^*}\vbeta\bigr)
,
\]
where $\mX$ and $\mX_{\vx^*}$ are matrices of covariates
(``design matrices'') for locations $\vfx$ and $\vx^*$,
respectively, and the other notations are analogous to those
in~(\ref{eq:PP-formulation}).
The matrices $\mQ_{\fY, \vY^*}$ and $\mQ_{\vY^*, \vY^*}$
are determined by covariance parameters $\vtheta$.
The parameters of this formulation have three components:
trend coefficients $\vbeta$, covariance parameters $\vtheta$,
and pilot point values $\vy^*$.
An objective function analogous to that of the first optimization
in the quasi-linear method is defined,
involving data reproduction, regularization, and functions of the
covariance parameters.
For each of a finite set of
$(\vbeta, \vtheta)$ values,
the objective function is optimized with respect to the pilot point
values.
The best $(\vbeta, \vtheta)$ value is chosen using empirical criteria
and the terminating values of the objective function.
The corresponding terminating pilot point values are taken as estimates
of $\vy^*$.
Subsequently,
the conditional mean of the field is as in the formula above
with $\vbeta$, $\vtheta$, and $\vy^*$ taking their estimated values,
and the conditional covariance is taken to be
$\mQ_{\fY,\fY} -
 \mQ_{\fY,\vY^*} \mQ_{\vY^*,\vY^*}^{-1} \mQ_{\vY^*,\fY}$.
Subsequent generation of field realizations
entails random draws from the normal distribution defined by
these conditional mean and covariance matrix.

These two methods share with anchored inversion
two important advantages over the pilot point method.
First, the structural parameters are estimated taking into account the
data $\vzb$.
Second, generation of realizations from the estimated normal
distribution is essentially free
(if the quasi-linear procedure foregoes the second optimization step).

The pilot points in \citet{Hernandez:2006:ISM} are equivalent to
inverted anchor points in anchored inversion.
As with the pilot point method,
the pilot point parameterization may be generalized to any linear
functions of the field, just as in anchored inversion.
The quasi-linear method has the effect of defining anchors
$\mH(\vmu) \fY$.
This implicit definition differs from anchored inversion in that
(1) the number of these anchors equals the length of the data vector
$\vzb$, ruling out the option to use more or fewer anchors;
(2) although these anchors are very effective,
they are unknown until the mean field $\vmu$ has been estimated.

In these two methods, the mean and variance
of the Gaussian model for the field have one set of optimal estimates.
This differs fundamentally from anchored inversion,
in which the Gaussian model parameters follow a multivariate posterior
distribution.


\subsection{Difficulties in a common formulation of the likelihood with respect to
type-B data}
\label{sec:common-likelihood-formulation}

The pilot point method and other methods discussed above
make use of the construct
\[
\bigl(\fM(\vfy) - \vzb\bigr)^T
\mW^{-1}
\bigl(\fM(\vfy) - \vzb\bigr)
,
\]
which is legitimate for least squares estimation or optimization
as long as the matrix $\mW$ is positive definite.
In some studies in the literature of stochastic hydrogeology
\citep{McLaughlin:1996:RGI, Kowalsky:2004:EFP, Li:2007:TDC},
this construct is expanded to a normal ``likelihood'' of
$\vfy$ with respect to $\vzb$:
\begin{equation}\label{eq:common-likelihood-formulation}
(2\pi)^{-n/2}
|\mW|^{-1/2}
\exp\Bigl\{-\frac{1}{2}
    \bigl(\fM(\vfy) - \vzb\bigr)^T
    \mW^{-1}
    \bigl(\fM(\vfy) - \vzb\bigr)
    \Bigr\}
,
\end{equation}
where $n$ is the length of the data vector $\vzb$.
We notice that this interpretation has two requirements.
First,
$\fM(\vfy) - \vzb$ is the observed value of a random variable
conditional on the model parameters, which is $\vfy$ here.
Second,
this random variable indeed has a normal distribution with mean zero and
covariance $\mW$.
We have established in Section~\ref{sec:errors}
that this random variable is ${\verrb}_1 - {\verrb}_2$,
where ${\verrb}_1$ is model error and ${\verrb}_2$ is measurement error.
We shall examine the two requirements in view of this fact.

The first requirement suggests that errors,
either in measurement or in model or in both,
must exist for this formulation to be meaningful.
This has un-natural effects in synthetic studies.
In probably all synthetic studies that we have read,
model error is absent, \ie,
the same forward computer code is used for generating synthetic data and
in the inversion operations.
This necessitates introducing noise to the synthetic data in order to
use the likelihood formulation above.
The \emph{known} covariance matrix of the synthetic noise,
$\mW$, is critical information for the construction
of~(\ref{eq:common-likelihood-formulation}), and this covariance
is not subject to ``estimation''.
In \citet{Hernandez:2006:ISM},
the ``measurements are taken to be error free'' (p.~12)
yet a variance of the measurement error is estimated.
On another occasion in that article (p.~7),
random noise is added to the synthetic data and the estimated error
variance is close to the true variance of the added noise.
However, having this quantity subject to estimation in data fitting is
questionable in the first place (if one is to maintain a likelihood
interpretation), and it is unclear why its estimate
turns out to be (in their method), or should be, close to the truth.
Similar inconsistencies have appeared in other studies.

Now comes the second requirement: how to specify the matrix $\mW$.
Despite the fact that the random variable involved is the combination of
measurement and model errors,
some studies have called it ``measurement error''.
This nomenclature can mislead the user to
specify its magnitude (or distribution) with only the
measurement mechanism in mind and assume independence between its components.
\citet{Li:2007:TDC} call it ``epistemic error''
and include in it error in intermediate models that are used to obtain the
observation $\vzb$.
This is an improved view over purely ``measurement'' error,
but still ignores error in the forward model.

The biggest concern over the
formulation~(\ref{eq:common-likelihood-formulation}) is that
the error covariance matrix ($\mW$) is hardly known to any high
level of accuracy
(see Section~\ref{sec:errors} for complications in the model error),
yet it is a center piece in the evaluation of the
likelihood.
The likelihood becomes more and more sensitive to the specification of
$\mW$ as the data and model quality increases.
In the limit where both the forward model and the measurement are free
of error, which is perfectly legitimate in synthetic studies,
the formulation simply fails.

This awkward situation has been noticed in
\citet{Kitanidis:1995:QLG}, which observes that the quasi-linear method
proposed in that article performs poorly
when the measurements ``have very small observation errors'' (p.~2418).

There have been mentions to this issue in the literature.
\citet{Lee:2002:MRF} recognize the model error component,
and find it a delicate task to specify $\mW$.
They take an empirical approach to this specification.
Also see the references cited by \citet{Lee:2002:MRF}.
\citet{Scales:2001:PIU} clearly state that the random variable
behind $\fM(\vfy) - \vzb$ is
the sum of measurement error and model error.
Questions are also raised by \citet[sec.~3.1]{Ginn:1990:IMS}
regarding the likelihood
formulation~(\ref{eq:common-likelihood-formulation}).
\citet{Tarantola:1982:IPQ} make it one of the minimum requirements that
the formulation of an inverse problem ``must be general enough to
incorporate theoretical errors in a natural way.''
The ``theoretical error'' in their terminology is equivalent to the
``model error'' here.
They give a telling example that,
``in seismology, the theoretical error made by solving the forward
travel time problem is often one order of magnitude larger than the
experimental error of reading the arrival time on a seismogram.''

As is explained in Section~\ref{sec:errors},
the proposed anchored inversion avoids this conceptual
(and consequently practical) difficulty by making the relation between
observations ($\vzb$) and model parameters ($\vTheta$) a well defined
statistical one, regardless of the existence or absence of ``errors''.
It is our hope that the preceding paragraphs have clarified
a long, widely overlooked issue,
and have suggested feasible directions for effort:
either quantify the model error, or render it negligible.
We believe it is very important in applications to
recognize that the forward model is almost never perfect,
yet method testing in synthetic settings often assume the model is
perfect.

\section{Summary}
\label{sec:summary}

\subsection{Contribution of the paper}

We proposed a general method,
named ``anchored inversion'',
for modeling spatial Gaussian processes (random fields)
with an emphasis on assimilating
various kinds of data in a systematic procedure.
The method is based on a new parameterization device
named ``anchors'' and a data classification scheme.
Anchors relate to linear functions of the field.
This device is useful due to a basic property of Gaussian processes,
namely, a Gaussian process conditional on a linear function of the field
is still a Gaussian process, with conditional mean and (co)variance
known analytically.
This makes anchors viable parameters for describing the random field
and providing a powerful mechanism to introduce (or impose) flexible
global and local features.
The data classification scheme is based on an abstraction of
the data-unknown relation.
Since the usage of a particular dataset is based on its position in this
classification,
data of radically different natures can be used
simultaneously as long as reliable (numerical) models exist for the
data-generation processes.
The method is blind to disciplinary details and technical
implementations of the data-generation processes.

According to the data classification, a dataset is of either type-A
or type-B.
Informally speaking,
type-A data are linear functions of the field,
whereas type-B data are nonlinear functions of the field.
The inverse procedure transfers information in nonlinear data
into the form of linear data, thus transforms an unfamiliar problem to a
familiar one.
The information transfer is in a statistical sense;
it can not be complete or exact.
The benefit is rigor and relative ease in statistical inference
as well as analysis of the results.

The formulation of the method
leads to a radical reduction in the dimensionality of the model
parameters in comparison with the common state-space approach
that uses the entire spatial field as a model parameter vector.
Benefit of this dimension reduction is, again,
rigor and relative ease in statistical inference and result analysis.

The extremely high dimensionality of model parameters
(in an state-space approach) has long troubled modelers.
This contributes to the ``identifiability'' or ``ill-posedness'' issue,
that is, there is no unique, best solution to the problem;
many solutions are equally good by one's chosen criterion.
This phenomenon is termed ``equifinality'' by
\citet{Beven:2006:MET},
who identifies as an important future research area
``how to use model dimensionality reduction to reduce the potential for
equifinality.''
The ill-posedness implies that, if tackled from a statistical
perspective,
the model parameter vector has no maximum likelihood estimate
\citep{OSullivan:1986:SPI}.
With radically reduced parameter dimensionality in the proposed
anchored inversion,
the problem could become well-posed \emph{in terms of the much shorter
parameter vector}.

Moreover,
compared to a state-space approach,
one has complete control over the number and choice of anchor
parameters.
The parameter dimensionality is no longer ``hijacked''
by the numerical model grid,
and the resolution of the model grid is no long restricted
by difficulties with parameter dimensionality.
These two things,
which are better to be separable,
are now separate.


Anchored inversion is not centered at optimization.
This has conceptual as well as computational implications.
Main conceptual advantages are
(1) statistical interpretations of the result
(see comments in Section~\ref{sec:connections}),
and
(2) avoidance of difficult technical issues including
defining an objective function (especially relative weights of different
data components and terms)
and choosing a termination criterion.
The main computational advantage is that
the cost of generating multiple field realizations by anchored inversion
is not determined by the cost of an optimization algorithm that is
performed for each realization.

We proposed a sampling-based algorithm that approximates the parameter
posterior distribution by a normal mixture.
The algorithm requires evaluation of the forward model exactly once
for each value examined for the model parameter vector.
The algorithm is quite general and is well suited to the problem of
conditioning on nonlinear data.
After inference of the model parameters,
realizations of the spatial field can be created at negligible
computational cost.

The proposed anchored inversion has a modular structure that is worth noting.
The method makes little assumption about
details of the structural parameters and their prior specification.
The choice of anchor parameters is a separate task,
as is development of the forward model.
Even the sampling method described in Section~\ref{sec:sampling} is
a modular component, although we expect the ideas of
normal mixture approximation and conditioning
to remain important efficiency features in future refinement of
this computationally demanding component.

We commented on connections of anchored inversion to existing methods
in stochastic hydrogeology.
Special emphasis was placed on a conceptual issue that involves
treatment of measurement and model errors.
We pointed out deficiencies related to this issue
in some state-space approaches, and explained how anchored inversion
tackles these difficulties.

\subsection{Future work}

The most obvious omission in this paper is how to choose anchor
parameters, including the number of these parameters and the exact
definition of each. In the case of anchor points,
the definition means the locations of anchors.

The proposed algorithm for sampling the parameter posterior needs to be
refined, possibly both in depth and in breadth.
In depth, it may be possible to develop an iterative procedure
to achieve sequential refinement.
In breadth, it may be useful to run multiple independent ``threads'' of
the algorithm in parallel, comparing/merging results and monitoring for
convergence.
Technical details of the normal mixture (kernel density estimation),
in particular determination of bandwidth (the scaling
factor $h$ in~(\ref{eq:joint-normal-mixture}))
need explorations
\citep[see][]{Jones:1996:BSB}.

The proposed method makes little assumption about the parameterization
of the random field by structural parameters $\vtheta$;
neither does not restrict the specification of a prior for $\vtheta$.
We illustrated with a simple formulation (Appendix~B).
A particular application may warrant more sophisticated formulations
such as anisotropy and
flexible spatial correlation models, especially the Mat\'ern model
\citep{Stein:1999:ISD, Banerjee:2003:SPS, Marchant:2007:MVM, Zhang:2007:ACM}.

\section*{Appendix A: properties of the multivariate normal
distribution}

The following are standard results;
see, \eg, \citet[chap~4]{Johnson:1998:AMS}.

If
$\vec{X} \sim \gauss(\vmu, \mSigma)$, then
\begin{equation}\tag{A-1}
\mH\vec{X} \sim \gauss(\mH\vmu, \mH\mSigma\mH^T)
.
\end{equation}

Let
$\vec{X} = \begin{bmatrix}\vec{X}_1 \\ \vec{X}_2\end{bmatrix}$
be distributed as $\gauss(\vmu, \mSigma)$ with
$\vmu = \begin{bmatrix}\vmu_1 \\ \vmu_2\end{bmatrix}$
and
$\mSigma = \begin{bmatrix}
    \mSigma_{11} & \mSigma_{12} \\
    \mSigma_{21} & \mSigma_{22}\end{bmatrix}$,
and $|\mSigma_{22}| > 0$.
Then given $\vec{X}_2 = \vx_2$,
the conditional distribution of $\vec{X}_1$ is
\begin{equation}\tag{A-2}
\vec{X}_1 \given \vx_2
\sim
\gauss\bigl(
    \vmu_1 + \mSigma_{12} \mSigma_{22}^{-1} (\vx_2 - \vmu_2)
    ,\;
    \mSigma_{11} - \mSigma_{12} \mSigma_{22}^{-1} \mSigma_{21}
    \bigr)
.
\end{equation}

If we have
$\vec{X} \sim \gauss(\vmu, \mSigma)$
and
$\vec{Y} = \mat{H}\vec{X}$, then from (A-1),
\begin{equation}\tag{A-3}
\begin{bmatrix}\vec{X}\\ \vec{Y}\end{bmatrix}
=
    \begin{bmatrix}\mat{I}\\ \mat{H}\end{bmatrix} \vec{X}
\sim \gauss\left(
    \begin{bmatrix}\vmu\\ \mat{H}\vmu\end{bmatrix}
    ,\;
    \begin{bmatrix}
        \mSigma & \mSigma \mH^T\\
        \mH\mSigma & \mH\mSigma\mH^T
        \end{bmatrix}
    \right)
.
\end{equation}
Given $\vec{Y} = \vec{y}$, then (A-2) and (A-3) suggest
the following conditional distribution,
which is convenient for the current study:
\begin{equation}\tag{A-4}
\vec{X} \given \vec{y}
\sim \gauss\bigl(
    \vmu
    + \mSigma\mH^T\bigl(\mH\mSigma\mH^T\bigr)^{-1}
        \bigl(\vec{y} - \mH\vmu\bigr)
    ,\;
    \mSigma
    - \mSigma\mH^T\bigl(\mH\mSigma\mH^T\bigr)^{-1}
        \mH\mSigma
    \bigr)
.
\end{equation}

\section*{Appendix B: structural parameters, prior specification,
posterior with respect to direct measurements, and sampling}

\newcommand\corrpar{\vec{\phi}}
\newcommand\scale{\varphi}
\newcommand\quadratic[2]{\ensuremath{\lVert #1\rVert_{#2}^2}}
\newcommand\abs[1]{\lvert #1 \rvert}

We model the spatial process $\Y$ as the composition of two components:
a mean function describing the expected value of $\Y(\x)$,
and a stochastic process describing fluctuations around the mean.
The random vector $\vec{Y}(\vx)$ is assumed to have a multinormal distribution:
\begin{equation}\label{eq:my-spatial-model}\tag{B-1}
\vec{Y}(\vx)
\sim \gauss\bigl(\mX\vbeta, \eta^2\mR\bigr)
,
\end{equation}
where
$\mX$ is a ``design matrix'' of known covariates (with a leading column of 1's),
$\vbeta$ is a vector of trend coefficients,
$\eta^2$ is the variance,
and
$\mR$ is the correlation matrix of $\vec{Y}$ at locations $\vx$.
The correlation between $\Y$'s at locations
$\x_1$ and $\x_2$ is modeled as
$\rho(\x_1, \x_2; \corrpar)$
with parameter vector $\corrpar$.
The vector $\corrpar$ usually contains components
for the range of correlation, smoothness of the spatial process,
nugget effects, and anisotropy.
In this formulation,
the structural parameter vector is
$\vtheta = (\corrpar, \eta^2, \vbeta)$.

We use the following prior:
\begin{equation}\label{eq:my-structural-prior}\tag{B-2}
p(\vtheta)
= p(\vbeta, \eta^2 \given \corrpar) \cdot p(\corrpar)
= p(\corrpar) / (\eta^2)^a
,
\end{equation}
where $a$ is a constant, taken to be 1 in this study.
The ``noninformative'' prior for $\vbeta$ and $\eta^2$
conditional on $\corrpar$ is a typical choice;
see \citet[secs 1.3, 2.2--2.4, 8.2]{Box:1973:BIS}
and \citet{Berger:2001:OBA}.
The prior for the correlation parameters $\corrpar$ is left flexible
and carried along in the analytical derivation;
the specific choice for $p(\corrpar)$ in this study is
uniform in a bounded interval (see below).

With the model~(\ref{eq:my-spatial-model})
and the prior~(\ref{eq:my-structural-prior}),
given point values $(\vx_*, \vy_*)$,
we can derive the posterior in a hierarchical fashion as follows:
\begin{align}
\vbeta \given \eta^2, \corrpar, \vy_*
&\;\sim\;
\gauss\bigl(
    \mQ^{-1} \mX_*^T \mR_*^{-1} \vy_*,\;
    \eta^2\, \mQ^{-1}
    \bigr)
\tag{B-3}\label{eq:post-trend}
\\
\eta^2 \given \corrpar, \vy_*
&\;\sim\;
\text{Inv-}\chi^2\Bigl(
    n + 2a - d_\beta - 2,\;
    (n + 2a - d_\beta - 2)^{-1}\, S^2\Bigr)
\tag{B-4}\label{eq:post-var}
\\
p(\corrpar \given \vy_*)
&\;\propto\;
  p(\corrpar)\,
  \abs{\mR_*}^{-1/2}\,
  \abs{\mQ}^{-1/2}\,
  \bigl(S^2\bigr)^{-(n + 2a - d_\beta - 2)/2}
\tag{B-5}\label{eq:post-corr}
\end{align}
where
$n$ is the length of data $\vy_*$,
$n_\beta$ is the dimension of $\vbeta$,
$\mX_*$ is the design matrix for the data locations $\vx_*$,
$\mR_*$ is the correlation matrix of $\vY$ at the data locations $\vx_*$,
$\mQ = \mX_*^T \mR_*^{-1} \mX_*$,
and
$
S^2
= \vy_*^T \,
  \bigl(
      \mR_*^{-1}
      - \mR_*^{-1} \mX_* \mQ^{-1} \mX_*^T \mR_*^{-1}
  \bigr)
  \,
  \vy_*
$.
The symbols $\gauss$ and $\text{Inv-}\chi^2$ stand for the normal
and scaled inverse-chi-square distributions, respectively.
Derivations of these results can be found in
\citet{Kitanidis:1986:PUE},
\citet[secs~2.6--2.7, 3.2--3.4, 3.6]{Gelman:1995:BDA},
and
\citet[chap.~7]{Diggle:2007:MBG}.

For illustrations in this study,
we use the isotropic exponential correlation function without nugget,
that is,
\[
\rho(x_1, x_2; \corrpar)
= \exp(-|x_1 - x_2| / \lambda)
,
\]
where $|\cdot|$ is the Euclidean distance
and $\lambda > 0$ is a range parameter,
which is the only component of $\corrpar$.
The prior of the range parameter $\lambda$ is taken to be uniform
on $(0.05L, L)$, where $L$ is the size of the model domain.
While this support of $\lambda$ is chosen subjectively,
there is no need to consider much larger ranges,
because correlation at long distances is better captured in the mean function
\citep[see][sec.~5.2.5]{Diggle:2007:MBG},
and because the variance and range parameters have an intimate
interaction.
According to \citet{deOliveira:1997:BPT},
a bounded support for the correlation parameter guarantees
a proper posterior.

To sample from the posterior distribution
$p(\vbeta, \eta^2, \corrpar \given \vy_*)$,
we first discretize the correlation parameters $\corrpar$
(here it is just a scalar $\lambda$) and pre-compute
$p(\corrpar \given \vy_*)$ for the set of $\corrpar$ values
according to~(\ref{eq:post-corr}),
then sample $\corrpar$ from this discretized set with weights
proportional to their posterior densities.
Given a sampled $\corrpar_*$,
we sample $\eta^2_*$ from the scaled inverse-chi-square distribution
$p(\eta^2 \given \corrpar_*, \vy_*)$ in~(\ref{eq:post-var}),
and finally sample $\vbeta_*$ from the normal distribution
$p(\vbeta \given \eta^2_*, \corrpar_*, \vy_*)$
in~(\ref{eq:post-trend}).

\newpage

\begin{figure}
\centering
\includegraphics{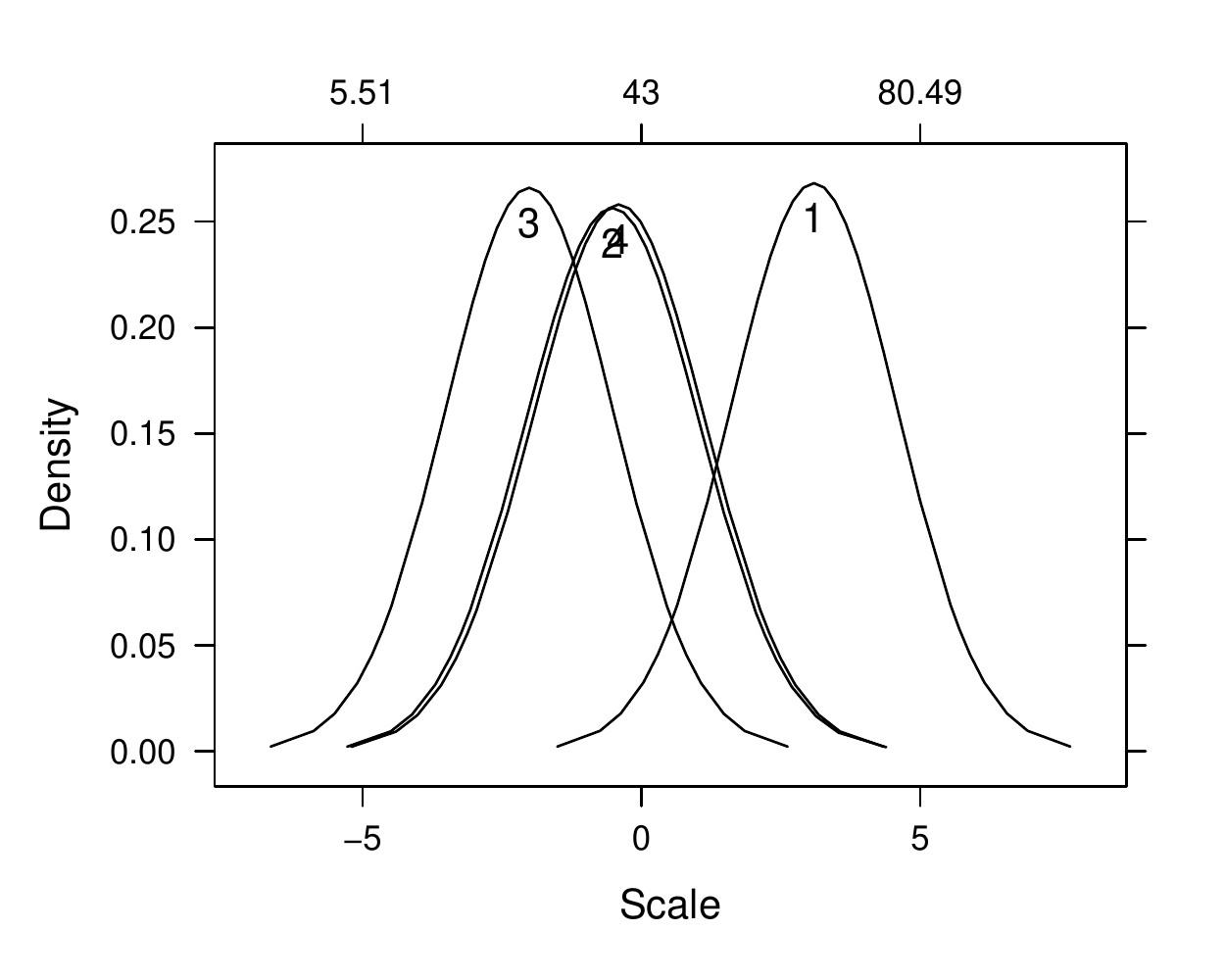}
\caption{Posterior distribution of the transformed scale parameter.
(Labels on the top are the back-transformed parameter values.)
Curves 1--4 correspond to sample sizes 5000, 10000, 20000, and 40000,
respectively.}
\label{fig:scale}
\end{figure}

\begin{thebibliography}{51}
\providecommand{\natexlab}[1]{#1}
\expandafter\ifx\csname urlstyle\endcsname\relax
  \providecommand{\doi}[1]{doi:\discretionary{}{}{}#1}\else
  \providecommand{\doi}{doi:\discretionary{}{}{}\begingroup
  \urlstyle{rm}\Url}\fi

\bibitem[{\textit{Banerjee and Gelfand}(2003)}]{Banerjee:2003:SPS}
Banerjee, S., and A.~E. Gelfand (2003), On smoothness properties of spatial
  processes, \textit{J. Multivariate Anal.}, \textit{84}, 85--100.

\bibitem[{\textit{Berger et~al.}(2001)\textit{Berger, {De Oliveira}, and
  Sans{\'o}}}]{Berger:2001:OBA}
Berger, J.~O., V.~{De Oliveira}, and B.~Sans{\'o} (2001), Objective {Bayesian}
  analysis of spatially correlated data, \textit{J. Am. Stat. Assoc.},
  \textit{96}(456), 1361--1374.

\bibitem[{\textit{Beven}(2006)}]{Beven:2006:MET}
Beven, K. (2006), A manifesto for the equifinality thesis, \textit{J. Hydrol.},
  \textit{320}, 18--36.

\bibitem[{\textit{Bindlish and Barros}(2002)}]{Bindlish:2002:SVR}
Bindlish, R., and A.~P. Barros (2002), Subpixel variability of remotely sensed
  soil moisture: an inter-comparison study of {SAR} and {ESTAR}, \textit{IEEE
  T. Geosci. Remote.}, \textit{40}(2), 326--337.

\bibitem[{\textit{Box and Tiao}(1973)}]{Box:1973:BIS}
Box, G. E.~P., and G.~C. Tiao (1973), \textit{Bayesian Inference in Statistical
  Analysis}, Addison-Wesley.

\bibitem[{\textit{Bube and Burridge}(1983)}]{Bube:1983:ODI}
Bube, K.~P., and R.~Burridge (1983), The one-dimensional inverse problem of
  reflection seismology, \textit{SIAM Review}, \textit{25}(4), 497--559.

\bibitem[{\textit{de~Marsily et~al.}(1984)\textit{de~Marsily, Lavedan, Boucher,
  and Fasanino}}]{deMarsily:1984:IIT}
de~Marsily, G., C.~Lavedan, M.~Boucher, and G.~Fasanino (1984), Interpretation
  of interference tests in a well field using geostatistical technniques to fit
  the permeability distribution in a reservoir model, in \textit{Geostatistics
  for Nattural Resources Characterization}, edited by G.~Verley, M.~David,
  A.~G. Journel, and A.~Marechal, {NATO ASI Ser. Series C}, pp. 831--849.

\bibitem[{\textit{{De Oliveira} et~al.}(1997)\textit{{De Oliveira}, Kedem, and
  Short}}]{deOliveira:1997:BPT}
{De Oliveira}, V., B.~Kedem, and D.~A. Short (1997), Bayesian prediction of
  transformed {Gaussian} random fields, \textit{J. Am. Stat. Assoc.},
  \textit{92}(440), 1422--1433.

\bibitem[{\textit{Diggle and {Ribeiro Jr.}}(2007)}]{Diggle:2007:MBG}
Diggle, P.~J., and P.~J. {Ribeiro Jr.} (2007), \textit{Model-based
  Geostatistics}, Springer.

\bibitem[{\textit{Doherty}(2003)}]{Doherty:2003:GWM}
Doherty, J. (2003), Ground water model calibration using pilot points and
  regularization, \textit{41}(2), 170--177.

\bibitem[{\textit{Evensen}(2003)}]{Evensen:2003:EKF}
Evensen, G. (2003), The {Ensemble Kalman Filter}: theoretical formulation and
  practical implementation, \textit{Ocean Dynamics}, \textit{53}, 343--367,
  \doi{10.1007/s10236-003-0036-9}.

\bibitem[{\textit{Fuentes and Raftery}(2005)}]{Fuentes:2005:MES}
Fuentes, M., and A.~E. Raftery (2005), Model evaluation and spatial
  interpolation by {Bayesian} combination of observations with outputs from
  numerical models, \textit{Biometrics}, \textit{61}, 36--45.

\bibitem[{\textit{Gelman et~al.}(1995)\textit{Gelman, Carlin, Stern, and
  Rubin}}]{Gelman:1995:BDA}
Gelman, A., J.~B. Carlin, H.~S. Stern, and D.~B. Rubin (1995), \textit{Bayesian
  Data Analysis}, Chapman \& Hall/CRC.

\bibitem[{\textit{Ginn and Cushman}(1990)}]{Ginn:1990:IMS}
Ginn, T.~R., and J.~H. Cushman (1990), Inverse methods for subsurface flow: a
  critical review of stochastic techniques, \textit{Stoch. Hydrol. Hydraul.},
  (4), 1--26.

\bibitem[{\textit{Givens and Raftery}(1996)}]{Givens:1996:LAI}
Givens, G.~H., and A.~E. Raftery (1996), Local adaptive importance sampling for
  multivariate densities with strong nonlinear relationships, \textit{J. Am.
  Stat. Assoc.}, \textit{91}(433), 132--141.

\bibitem[{\textit{G{\'o}mez-Hern{\'a}nez
  et~al.}(1997)\textit{G{\'o}mez-Hern{\'a}nez, Sahuquillo, and
  Capilla}}]{Gomez-Hernanez:1997:SST1}
G{\'o}mez-Hern{\'a}nez, J.~J., A.~Sahuquillo, and J.~E. Capilla (1997),
  Stochastic simulation of transmissivity fields conditional to both
  transmissivity and piezometric data--{I}.\ theory, \textit{J. Hydrol.},
  \textit{203}, 162--174.

\bibitem[{\textit{Hernandez et~al.}(2006)\textit{Hernandez, Neuman, Guadagnini,
  and Carrera}}]{Hernandez:2006:ISM}
Hernandez, A.~F., S.~P. Neuman, A.~Guadagnini, and J.~Carrera (2006), Inverse
  stochastic moment analysis of steady state flow in randomly heterogeneous
  media, \textit{Water Resour. Res.}, \textit{42}, W05425,
  \doi{10.1029/2005WR004449}.

\bibitem[{\textit{Johnson and Wichern}(1998)}]{Johnson:1998:AMS}
Johnson, R.~A., and D.~W. Wichern (1998), \textit{Applied Multivariate
  Statistial Analysis}, 4th ed., Prentice Hall.

\bibitem[{\textit{Jones et~al.}(1996)\textit{Jones, Marron, and
  Sheather}}]{Jones:1996:BSB}
Jones, M.~C., J.~S. Marron, and S.~J. Sheather (1996), A brief survey of
  bandwidth selection for density estimation, \textit{J. Am. Stat. Assoc.},
  \textit{91}(433).

\bibitem[{\textit{Kass and Wasserman}(1996)}]{Kass:1996:SPD}
Kass, R.~E., and L.~Wasserman (1996), The selection of prior distribution by
  formal rules, \textit{J. Am. Stat. Assoc.}, \textit{91}(435), 1343--1370.

\bibitem[{\textit{Kitanidis}(1986)}]{Kitanidis:1986:PUE}
Kitanidis, P.~K. (1986), Parameter uncertainty in estimation of spatial
  functions: {Bayesian} analysis, \textit{Water Resour. Res.}, \textit{22}(4),
  499--507.

\bibitem[{\textit{Kitanidis}(1995)}]{Kitanidis:1995:QLG}
Kitanidis, P.~K. (1995), Quasi-linear geostatistical theor for inversing,
  \textit{Water Resour. Res.}, (10), 2411--2419.

\bibitem[{\textit{Kowalsky et~al.}(2004)\textit{Kowalsky, Finsterle, and
  Rubin}}]{Kowalsky:2004:EFP}
Kowalsky, M.~B., S.~Finsterle, and Y.~Rubin (2004), Estimating flow parameter
  distributions using ground-penetrating radar and hydrological measurements
  during transient flow in the vadose zone, \textit{Adv. Water Resour.},
  \textit{27}, 583--599.

\bibitem[{\textit{Lee et~al.}(2002)\textit{Lee, Higdon, Bi, Ferreira, and
  West}}]{Lee:2002:MRF}
Lee, H. K.~H., D.~M. Higdon, Z.~Bi, M.~A.~R. Ferreira, and M.~West (2002),
  {Markov} random field models for high-dimensional parameters in simultaitons
  of fluid flow in porous media, \textit{Technometrics}, \textit{44}(3),
  230--241, \doi{10.1198/004017002188618419}.

\bibitem[{\textit{Li et~al.}(2007)\textit{Li, Englert, Cirpka, Vanderborght,
  and Vereecken}}]{Li:2007:TDC}
Li, W., A.~Englert, O.~A. Cirpka, J.~Vanderborght, and H.~Vereecken (2007),
  Two-dimensional characterization of hydraulic heterogeneity by multiple
  pumping tests, \textit{Water Resour. Res.}, \textit{43}, W04433,
  \doi{10.1029/2006WR005333}.

\bibitem[{\textit{Marchant and Lark}(2007)}]{Marchant:2007:MVM}
Marchant, B.~P., and R.~M. Lark (2007), The {Mat\'ern} variogram model:
  {Implications} for uncertainty propagation and sampling in geostatistical
  surveys, \textit{Geoderma}, \textit{140}, 337--345,
  \doi{10.1016/j.geoderma.2007.04.016}.

\bibitem[{\textit{Marjoram et~al.}(2003)\textit{Marjoram, Molitor, Plagnol, and
  Tavar\'e}}]{Marjoram:2003:MCM}
Marjoram, P., J.~Molitor, V.~Plagnol, and S.~Tavar\'e (2003), {Markov} chain
  {Monte} {Carlo} without likelihoods, \textit{Proceedings of the National
  Academy of Sciences}, \textit{100}, 15,324--15,328,
  \doi{10.1073/pnas.0306899100}.

\bibitem[{\textit{Marron and Wand}(1992)}]{Marron:1992:EMI}
Marron, J.~S., and M.~P. Wand (1992), Exact mean integrated squared error,
  \textit{Ann. Statist.}, \textit{20}(2), 712--736.

\bibitem[{\textit{{McLachlan}}(2000)}]{McLachlan:2000:FMM}
{McLachlan}, G. (2000), \textit{Finite Mixture Models}, John Wiley \& Sons,
  Inc.

\bibitem[{\textit{{McLaughlin} and Townley}(1996)}]{McLaughlin:1996:RGI}
{McLaughlin}, D., and L.~R. Townley (1996), A reassessment of the groundwater
  inverse problem, \textit{Water Resour. Res.}, (5), 1131--1161.

\bibitem[{\textit{Merlin et~al.}(2005)\textit{Merlin, Chehbouni, Kerr, Njoku,
  and Entekhabi}}]{Merlin:2005:CMM}
Merlin, O., A.~G. Chehbouni, Y.~H. Kerr, E.~G. Njoku, and D.~Entekhabi (2005),
  A combined modeling and multispectral/multiresolution remote sensing approach
  for disaggregation of surface soil moisture: application to {SMOS}
  configuration, \textit{IEEE T. Geosci. Remote.}, \textit{43}(9).

\bibitem[{\textit{Newsam and Enting}(1988)}]{Newsam:1988:IPA}
Newsam, G.~N., and I.~G. Enting (1988), Inverse problems in atmospheric
  constituent studies: {I.} determination of surface sources under a diffusive
  transport approximation, \textit{Inverse Problems}, \textit{4}, 1037--1054.

\bibitem[{\textit{Nowak and Cirpka}(2004)}]{Nowak:2004:MLM}
Nowak, W., and O.~A. Cirpka (2004), A modified {Levenberg-Marquardt} algorithm
  for quasi-linear geostatistical inversin, \textit{Adv. Water Resour.}, pp.
  737--750.

\bibitem[{\textit{O'Sullivan}(1986)}]{OSullivan:1986:SPI}
O'Sullivan, F. (1986), A statistical perspective on ill-posed inverse problems,
  \textit{Statistical Science}, \textit{1}(4), 502--518.

\bibitem[{\textit{RamaRao and LaVenue}(1995)}]{RamaRao:1995:PPM1}
RamaRao, B.~S., and A.~M. LaVenue (1995), Pilot point methodology for automated
  calibration of an ensemble of conditionally simulated transmissivity fields
  1.\ theory and computational experiments, \textit{Water Resour. Res.},
  \textit{31}(3), 475--493.

\bibitem[{\textit{Robert and Casella}(2005)}]{Robert:2005:MCS}
Robert, C.~P., and G.~Casella (2005), \textit{Monte Carlo Statistical Methods},
  2nd ed., Springer.

\bibitem[{\textit{Rubin}(2003)}]{Rubin:2003:ASH}
Rubin, Y. (2003), \textit{Applied Stochastic Hydrogeology}, Oxford University
  Press.

\bibitem[{\textit{Sambridge and Mosegaard}(2002)}]{Sambridge:2002:MCM}
Sambridge, M., and K.~Mosegaard (2002), {Monte Carlo} methods in geophysical
  inverse problems, \textit{Reviews of Geophysics}, \textit{40}(3),
  \doi{10.1029/2000RG000089}.

\bibitem[{\textit{Scales and Tenorio}(2001)}]{Scales:2001:PIU}
Scales, J.~A., and L.~Tenorio (2001), Prior information and uncertainty in
  inverse problems, \textit{Geophysics}, \textit{66}(2), 389--397.

\bibitem[{\textit{Scott}(1992)}]{Scott:1992:MDE}
Scott, D.~W. (1992), \textit{Multivariate Density Estimation}, John Wiley \&
  Sons, Inc.

\bibitem[{\textit{Scott and Wand}(1991)}]{Scott:1991:FMD}
Scott, D.~W., and M.~P. Wand (1991), Feasibility of multivariate density
  estimates, \textit{Biometrika}, \textit{78}(1), 197--205.

\bibitem[{\textit{Sheather}(2004)}]{Sheather:2004:DE}
Sheather, S.~J. (2004), Density estimation, \textit{Statistical Science},
  \textit{19}(4), 588--597, \doi{10.1214/088342304000000297}.

\bibitem[{\textit{Stein}(1999)}]{Stein:1999:ISD}
Stein, M.~L. (1999), \textit{Interpolation of Spatial Data: Some Theory for
  Kriging}, Springer.

\bibitem[{\textit{Tarantola and Valette}(1982)}]{Tarantola:1982:IPQ}
Tarantola, A., and B.~Valette (1982), Inverse problems = question for
  information, \textit{Journal of Geophysics}, \textit{50}, 159--170.

\bibitem[{\textit{Tenorio}(2001)}]{Tenorio:2001:SRI}
Tenorio, L. (2001), Statistical regularization of inverse problems,
  \textit{SIAM Review}, \textit{43}(2), 347--366.

\bibitem[{\textit{Tikhonov and Arsenin}(1977)}]{Tikhonov:1977:SIP}
Tikhonov, A., and V.~Arsenin (1977), \textit{Solution of Ill-posed Problems},
  Winston, Washington, DC.

\bibitem[{\textit{Wand and Jones}(1995)}]{Wand:1995:KS}
Wand, M.~P., and M.~C. Jones (1995), \textit{Kernel Smoothing}, Chapman \&
  Hall/CRC.

\bibitem[{\textit{Welch and Bishop}(1995)}]{Welch:1995:IKF}
Welch, G., and G.~Bishop (1995), An introduction to the {Kalman} filter,
  \textit{Tech. Rep. TR 95-041}, University of North Carolina at Chapel Hill,
  updated in 2006.

\bibitem[{\textit{West}(1993)}]{West:1993:APD}
West, M. (1993), Approximating posterior distributions by mixture, \textit{J.
  R. Stat. Soc., B}, \textit{55}(2), 409--422.

\bibitem[{\textit{Yeh}(1986)}]{Yeh:1986:RPI}
Yeh, W. W.-G. (1986), Review of parameter identification procedures in
  groundwater hydrology: the inverse problem, \textit{Water Resour. Res.},
  \textit{22}(2), 95--108.

\bibitem[{\textit{Zhang et~al.}(2007)\textit{Zhang, Beletsky, Schwab, and
  Stein}}]{Zhang:2007:ACM}
Zhang, Z., D.~Beletsky, D.~J. Schwab, and M.~L. Stein (2007), Assimilation of
  current measurements into a circulation model of {Lake Michigan},
  \textit{Water Resour. Res.}, \textit{43}, W11407, \doi{10.1029/2006WR005818}.

\end{thebibliography}
\end{document}